\documentclass[twocolumn]{emulateapj}
\usepackage{amsmath}

\newcommand{\Msolar}{M$_{\odot}$}

\newcommand{\kms}{km s$^{-1}$}

\begin{document}

\title{Direct \textit{N}-Body Modeling of the Old Open Cluster NGC 188: A Detailed Comparison
of Theoretical and Observed Binary Star and Blue Straggler Populations\footnote{WIYN Open Cluster Study. LI.}}
\shorttitle{Direct \textit{N}-body Modeling of NGC 188}

\author{Aaron M.~Geller\footnote{Visiting Astronomer, Kitt Peak National Observatory, National Optical Astronomy Observatory, which is operated by the Association of Universities for Research in Astronomy (AURA) under cooperative agreement with the National Science Foundation.}}
\affil{Center for Interdisciplinary Exploration and Research in Astrophysics (CIERA) and Department of Physics and Astronomy, Northwestern University, 2145 Sheridan Rd, Evanston, IL 60208, USA}
\affil{Department of Astronomy, University of Wisconsin - Madison, WI 53706 USA}
\email{a-geller@northwestern.edu}

\author{Jarrod R.~Hurley}
\affil{Centre for Astrophysics and Supercomputing, Swinburne University of Technology, VIC 3122, Australia}

\author{Robert D.~Mathieu$^\dag$}
\affil{Department of Astronomy, University of Wisconsin - Madison, WI 53706 USA}

\shortauthors{Geller, Hurley \& Mathieu}

\begin{abstract}

Following on from a recently completed
radial-velocity survey of the old (7 Gyr) open cluster NGC 188 
in which we study in detail the solar-type hard binaries and blue stragglers of the cluster,
here we investigate the dynamical evolution of NGC 188 through a sophisticated $N$-body model.
Importantly, we employ the observed binary properties of the young (150 Myr) open cluster M35, where possible,
to guide our choices for parameters of the initial binary population.
We apply pre-main-sequence tidal circularization and a substantial increase to the main-sequence tidal circularization rate, 
both of which are necessary to match the observed tidal circularization periods in the literature, including that of NGC 188. 
At 7 Gyr the main-sequence solar-type hard-binary population in the model matches that of NGC 188 
in both binary frequency and distributions of orbital parameters.
This agreement between the model and observations is in a large part due to the similarities 
between the NGC 188 and M35 solar-type binaries.
Indeed, among the 7 Gyr main-sequence binaries in the model, only those with $P \gtrsim 1000$ days 
begin to show potentially observable evidence for modifications by dynamical encounters, 
even after 7 Gyr of evolution within the star cluster.  This emphasizes the importance 
of defining accurate initial conditions for star cluster models, which we propose is best 
accomplished through comparisons with observations of young open clusters like M35.
Furthermore, this finding suggests that observations of the present-day binaries
in even old open clusters can provide valuable information on their primordial binary populations.
However, despite the model's success at matching the observed solar-type main-sequence population, the model
underproduces blue stragglers and produces an overabundance of long-period 
circular main-sequence -- white-dwarf binaries as compared to the true cluster. 
We explore several potential solutions to the paucity of blue stragglers and conclude that
the model dramatically underproduces blue stragglers through mass-transfer processes.
We suggest that common-envelope evolution may have been incorrectly imposed on the progenitors 
of the spurious long-period circular main-sequence -- white-dwarf binaries,
which perhaps instead should have gone through stable mass transfer to create blue stragglers, 
thereby bringing both the number and binary frequency of the blue straggler population 
in the model into agreement with the true blue stragglers in NGC 188.
Thus, improvements in the physics of mass transfer and common envelope evolution employed in the model 
may in fact solve both discrepancies with the observations.
This project highlights the unique accessibility of open clusters to both 
comprehensive observational surveys and full-scale $N$-body simulations, both of which have only recently 
matured sufficiently to enable such a project, and 
underscores the importance of open clusters to the study of star cluster dynamics.
\end{abstract}

\keywords{(galaxy:) open clusters and associations: individual (NGC 188) - (stars:) binaries: spectroscopic - (stars:) blue stragglers -  (methods:) numerical}

\section{Introduction}

NGC 188 is one of the most well studied open clusters, and is a cornerstone of the WIYN Open Cluster Study
\citep[WOCS;][]{mat00}.  Its old age ($\sim$7 Gyr; \citealt{sar99}, \citealt{mei09}), rich binary population \citep{gel09,gel12} and
abundance of blue stragglers \citep[BSs;][]{mat09,gel11}, X-ray sources \citep{bel98,gon05}, and other exotic stars \citep[see e.g.,][]{gel08,gel09}
make NGC 188 particularly important for the study of star cluster evolution.  

Sophisticated $N$-body codes now incorporate stellar dynamics and stellar evolution self-consistently (e.g., \texttt{NBODY6}; \citealt{aar03}),
enabling models of rich open clusters, like NGC 188, that include up to a few $10^4$ single and multiple stars initially \citep[e.g.,][]{por01,hur05}.
Until recently, the observations needed to test many of the important parameters in these models, and particularly those of the binary population,
have been unavailable, due in part to the decade(s) of multi-epoch observations necessary for detecting and defining the binary stars.
Our comprehensive radial-velocity survey of the single and binary stars in NGC 188 \citep{gel08,gel09,mat09,gel11,gel12}
now provide this essential empirical guidance, permitting us to model NGC 188 with a high level of accuracy.

Perhaps the most detailed $N$-body study of a specific open cluster to date is that of \citet{hur05}.
They aimed to model the open cluster M67, and particularly to match the observed color-magnitude diagram (CMD), with a 
focus on the BS population. 
Hurley et al.~initialized their binary population so as to maximize the production rate of BSs, and also to attempt to recreate the general 
binary properties of the M67 BSs.
In doing so, they managed to nearly match the observed number of BSs in M67 at 4 Gyr in their simulation, and they reproduced the 
true cluster CMD quite closely.  

As both NGC 188 and M67 are old open clusters (at $\sim$7 and $\sim$4 Gyr, respectively)
and the binary and BS populations of both clusters are remarkably similar, in \citet{gel12} we compare this simulation to 
the binary and BS populations observed in NGC 188.  We find that the M67 simulation contains far too many short-period binaries
as compared to our observations of NGC 188, and that the binary orbital parameters and hard-binary (periods  $<10^4$ days)
frequency of the BSs in the simulation do not match those
of NGC 188 (or M67).  These discrepancies are in a large part the result of the unrealistic initial binary population that was 
heavily weighted towards short-period systems, inconsistent with observed binary populations in young open clusters, like M35 
\citep[e.g.][and see also \citealt{mei05}]{mat08}.

In this paper, we model NGC 188 with a primary goal of matching the observed solar-type binary population.
We employ the observed binary population of the young open cluster M35 (180 Myr) to guide our definition of the initial binary frequency and 
distributions of period and eccentricity. Thus here we create the first $N$-body model whose 
initial binary population is defined by such detailed observations of a young open cluster.
We first discussed the BSs created in this model in \citet{gel11}. Here we present the NGC 188 model in full detail.

In Section~\ref{method} we discuss the simulation method. 
Section~\ref{RVobs} provides a brief outline of the observations used in guiding the model, and 
Section~\ref{init} explains the initial conditions of the model, including a detailed discussion of the 
initial binary population.  In Section~\ref{SOBS} we analyze the 7 Gyr cluster CMD, mass and structure
in comparison to observations of NGC 188.  We then compare the binary population and 
BSs in the model at
7 Gyr to the observed binaries and BSs in NGC 188 (Sections~\ref{final_binary} and~\ref{final_BS}). 
In Section~\ref{final_trips} we analyze the dynamically formed triples within our model as compared
to observations of triples in open clusters and the Galactic field.
Finally in Section~\ref{discuss} we investigate the dynamical evolution of the main-sequence (MS) 
binaries and evaluate BS production within our $N$-body model, 
and we then provide our conclusions in Section~\ref{conclusion}.

\section{Simulation Method} \label{method}

We use the \texttt{NBODY6} code \citep{aar03} to model the dynamical evolution 
within the NGC 188 simulations.
Stellar and binary evolution are included in NBODY6 using the work of \citet{hur00} and \citet{hur02}.
Much of our method is identical to that of \citet{hur05}, with one distinction being
that they used the \texttt{NBODY4} code with GRAPE-6 computing hardware 
\citep{mak03} while we use the \texttt{NBODY6} code on the supercomputer at the 
Swinburne Centre for Astrophysics and Supercomputing (without a GRAPE board).  
The main difference between the two code versions is that \texttt{NBODY4}
interfaces with the GRAPE hardware for the force calculations while
in \texttt{NBODY6} this is done on a standard CPU or more recently on a 
graphics processing unit \citep[GPU;][]{nit12}.
To increase efficiency, \texttt{NBODY6} also has the option to compute force
contributions from distant particles less frequently than from
near neighbors \citep{aar99}.
We modified the \texttt{NBODY6} code to define the initial binary population and output format.
We choose to output stellar evolution parameters, binary orbital properties, positions and velocities for 
all stars in snapshot intervals of $\sim$30 Myr.

Individual stars are evolved according to the Single-Star Evolution algorithm \citep[SSE;][]{hur00}.  
This code rapidly models all phases of evolution from the zero-age main sequence (ZAMS) 
through remnant phases, for stars covering a mass range of 0.1 - 100 \Msolar~and a wide 
range in metallicity.  Mass loss through stellar winds is included.
Binary stars are evolved according to the Binary-Star Evolution algorithm \citep[BSE;][]{hur02}.
This code contains prescriptions for tidal circularization and synchronization, angular 
momentum loss mechanisms (e.g., magnetic braking and gravitational radiation), mass transfer from 
Roche lobe overflow (RLOF), accretion from stellar winds, common-envelope (CE) events, mergers, and 
direct collisions.  Thus the code contains numerous pathways to create anomalous stars, 
such as BSs, even without the aid of dynamical encounters within a star cluster.  The additional
considerations that may occur in a cluster environment, such as
perturbations to binary orbits, exchanges, etc., are modeled in \texttt{NBODY6} \citep[][and references therein]{aar03}.

\citet{hur02,hur05} explain the modeling of BSs in detail.  
As a goal of this paper is to closely compare the observed and simulated BS populations, we provide a short summary here.
BSs can form in the $N$-body model through three primary mechanisms, namely mass transfer during 
RLOF, physical stellar collisions, and mergers of two stars within a binary.  
BS formation through wind mass transfer is also possible in principle, but in practice this formation 
channel is very infrequent.  Therefore for this paper, we will use the term mass transfer to refer specifically 
to mass transfer from RLOF.

Mass transfer during RLOF is modeled in \texttt{NBODY6} according to \citet{hur02}.  In short, 
the stability of mass transfer is determined using the radius-mass exponents $\zeta$ defined
by \citet{web85}, which are used to distinguish between mass transfer that proceeds on a dynamical, 
thermal or nuclear time scale.  BSs created through mass transfer from RLOF derive from 
thermal, and sometimes nuclear mass transfer, and are formed when a
MS star accretes mass from either a MS (Case A), red giant (Case B) or asymptotic giant (Case C) donor.
Case A mass transfer leads to a merger, while Cases B and C generally result in a BS in a detached binary 
with a white dwarf (WD) companion.  
Dynamical mass transfer 
either leads to a CE episode, when mass is transferred from a giant, or a merger, when mass 
is transferred from a low-mass ($\lesssim 0.7$ \Msolar) MS star, neither of which will produce a BS at 7 Gyr.
In practice the criteria to determine whether a binary undergoing
RLOF will enter dynamical mass transfer or thermal/nuclear mass transfer 
(e.g., possibly creating a BS) is determined by the critical mass ratio, $q_c$, such that if 
$M_{donor}/M_{accretor} > q_c$ the binary undergoes dynamical mass transfer.  Both the $q_c$ formulae from 
\citet{hur02} and \citet{hje87} are options in the $N$-body code, and we have chosen to implement 
the latter.  We return to the discussion of CE evolution and the $q_c$ value in Section~\ref{discuss_lPcirc}.

If a MS star gains mass through a mass-transfer episode, merger or collision, it 
is presumed to remain on the MS but attain a higher luminosity and effective temperature.
The lifetime of the star is determined based on the fraction of unburned hydrogen in the core of the star, 
such that the fraction of MS lifetime remaining is directly 
proportional to the fraction of remaining unburned hydrogen in the core.

During thermal and nuclear mass transfer, the rejuvenation procedure is determined based on the structure 
of the accretor star's core.
If mass is transferred onto a star with a radiative core ($0.35$~\Msolar$ \leq M \leq 1.25$~\Msolar)
the fraction of unburned hydrogen is barely changed.  Therefore the fraction of its
MS lifetime that has elapsed is unaffected by the mass transfer, but the 
effective age of the star is less than that of other normal stars of similar mass.
If mass is transferred onto a star with a 
convective core ($M > 1.25$~\Msolar) or that is fully convective ($M < 0.35$~\Msolar), the core 
grows and mixes in unburned hydrogen.  
It is assumed that the amount of burnt hydrogen is 
unchanged, and that the core mass grows directly proportional to the mass of the star, which is increased during
mass transfer.  The increase in the fraction of unburned hydrogen in the core increases the remaining
fraction of MS lifetime.
If mass transfer causes a star to move across the radiative/convective-core boundary the treatment simply
switches from one core type to the other.  
After the mass-transfer phase, the total MS lifetime of this rejuvenated star (from $t=0$ in the simulation to the time when the star evolves off the MS)
for either stars with convective or radiative cores is equal to the total lifetime of a normal
star of that mass.  The rejuvenation process described here ensures that the fraction of MS lifetime remaining for 
this star is greater than that of a normal MS star of this mass 
(regardless of whether the core is radiative or convective).

If two MS stars undergo a merger or collision, the product is taken to be a new MS star where
the stellar material is fully mixed.  A BS formed in this manner
then evolves along the MS similarly to normal MS stars of the same mass until 10\% of the 
total hydrogen has been burnt, at which point the BS evolves off of the MS and follows the 
standard evolutionary sequence for a star of this mass.  As discussed in \citet{hur05} the assumption 
of complete mixing here is likely incorrect and may affect the lifetime and appearance of these stars.
We follow this method here in part to aid comparisons with previous models that use similar prescriptions.
Though correcting for this over simplification is highly desired, it is beyond the scope of this project.
We also note that, in the $N$-body code, collision and merger products are assumed to achieve thermal equilibrium rapidly, and these BSs follow
equilibrium models upon creation.  However, detailed studies of collision products show that such BSs may not be in 
thermal equilibrium, and may be more luminous and have shorter lifetimes than the equilibrium models used here \citep[e.g.][]{gle08}.
We will return to this point briefly in Section~\ref{discuss}.

The processes involved in forming a BS can cause the star to lie above the MS turnoff on
a CMD.  Observationally we identify BSs as being generally both brighter and bluer than 
the MS turnoff.  Therefore here we will also use this definition, and not include any collision, merger, or mass-transfer 
products that still remain below the MS turnoff.
This empirical identification procedure necessarily introduces some ambiguity between binaries found near the 
turnoff, as a normal equal-mass MS-MS binary could be up to 0.75 magnitudes brighter than 
the turnoff (without containing a BS).
Within the simulation we can also identify BSs as being more massive
than a normal MS star at the turnoff, while still maintaining the structure and evolutionary 
state of a MS star.  Therefore this ambiguity does not effect BS identification in the 
$N$-body model.

In order to reduce the stochastic effects present in $N$-body simulations, and especially those 
of BS production, we ran twenty simulations, and for much of the analysis we average the results
together.  For all twenty simulations, the initial stellar and binary parameters (e.g., positions, velocities, 
binary periods and eccentricities, etc.) are all drawn from the same respective distributions, but 
each simulation randomizes the parameters to produce a unique initial stellar population.
In this paper we will use the term ``simulation'' to refer to any individual $N$-body simulation (e.g., any one of these twenty realizations), 
and we will use the term NGC 188 ``model'' to refer to the combination of these twenty simulations.  
We discuss our method for combining these simulations in later sections.

\section{Observed Binary Star Populations} \label{RVobs}

Here we give a brief outline of the observed stellar samples in M35 used to define our initial 
binary population and in NGC 188 for comparison with the 7 Gyr binary and BS populations in the $N$-body model.
The M35 radial-velocity sample is discussed in detail in \citet{gel10}.
In short, our M35 sample includes MS solar-type stars within the 
magnitude range of $13.0 \leq V \leq 16.5$ (1.6 - 0.8 \Msolar) and within 30 arcminutes 
from the cluster center (7 pc in projection at a distance of 805 pc or $\sim$4 core radii). 
The cluster turnoff in M35 is at $V\sim$9.5 ($\sim$4 \Msolar), and therefore our sample includes 
only MS stars.
\citet{gel10} identify 360 single and 55 binary cluster members within this sample; 
39 of these binaries have orbital solutions. We use this binary population to define the initial binary 
frequency and distributions of binary orbital parameters for the NGC 188 simulations (Section~\ref{init_binary}).  

The NGC 188 radial-velocity sample is discussed in detail in \citet{gel08}.  
Our stellar sample spans a magnitude range of $10.5 \leq V \leq 16.5$ (1.1 - 0.9 \Msolar),
and extends to a projected radius of 17 pc ($\sim$13 core radii, or 30 arcminutes). 
The sample includes the brightest stars in the cluster down to about two 
magnitudes below the MS turnoff, and all NGC 188 BSs.
We identify 358 single and 129 binary cluster members within this sample;
85 of these binaries have orbital solutions.
The distributions of orbital parameters are discussed in detail in \citet{gel12}
and we also show a sample of these results in Section~\ref{final_binary}.

For both clusters we can detect binaries out to periods of order $10^4$ days.  
The hard-soft boundary \citep{heg75} in both of these clusters is at periods of order $\sim10^6$ days, and 
therefore all detected binaries in these clusters are hard binaries.
Here we will use the term ``hard binary'' to refer specifically to binaries within the period range that 
we can detect with our observations.
The observed binary frequencies and distributions of orbital parameters
for M35 and NGC 188 presented here have been corrected for incompleteness using the method of 
\citet{gel12}, and importantly, our completeness in binary detection and orbital solutions is roughly 
equivalent for both clusters.  

\section{Initial Conditions of the Simulations} \label{init}

\subsection{Initial Cluster Structure} \label{init_struct}

NGC 188 is observed to have a solar metallicity and an age of 7 Gyr \citep[][and see also \citealt{mei09} and \citealt{for07}]{sar99}.
Empirical estimates for the mass of NGC 188 range from about 1500~\Msolar~to~3800~\Msolar~ 
\citep{bon05,gel08,chu10}.  In order to retain this mass by 7 Gyr, we scale from the results of 
the \citet{hur05} simulation to account for mass loss through evaporation and stellar evolution processes.
We find a suitable initial mass of 23600~\Msolar~(39000 stars).  We choose the stellar 
masses according to a \citet{kro01} initial-mass function (IMF).

We distribute the stars in space according to a Plummer density profile.  
In theory the Plummer model extends to infinite radius; therefore in practice we apply a cut-off
at ten times the half-mass radius to avoid rare cases of stars placed at large distances in the initial distribution.
This initial Plummer 
density profile quickly evolves to resemble a King profile within our model (as in other 
$N$-body models, e.g. \citealt{hur03,hur05}).  
The initial velocities are isotropic and are generated based on the positions and the assumption of dynamical 
equilibrium.

The initial size scale of a simulation is defined by the initial half-mass radius, and
we aim to match the observed core and tidal radii of the true cluster at 7 Gyr.
\citet{bon05} find NGC 188 to have a tidal radius of 21$\pm$4 pc and a core radius of 1.3$\pm$0.1 pc,
while \citet{chu10} find a tidal radius of 34.4$\substack{+16 \\ -10}$~pc and a core radius of 
2.1$\substack{+0.9 \\ -0.6}$~pc (both using standard \citealt{kin62,kin66} model fits).  
We note that \citet{chu10} included only proper-motion members of the cluster in their analysis while
\citet{bon05} used 2MASS photometry without any kinematic membership data, which may account for 
the discrepancies between these two results.
For our purposes, we assume that the true core and tidal radii of NGC 188 fall
somewhere between these values. 

We ran a number of test simulations with different half-mass radii to attempt to estimate an appropriate
initial half-mass radius that will evolve to match the observed values described above.  
We find that using an initial half-mass radius of 4.6~pc results in a reasonable agreement with the observed
core and tidal radii for NGC 188 at the age of 7 Gyr, see Section~\ref{SOBS}, and we use this initial 
half-mass radius for our NGC 188 model. 
(Also note that \citet{hur05} used an initial half-mass radius of 3.9~pc, and found 
a resulting core radius of 0.64 pc at 4 Gyr, much smaller than NGC 188.)

\citet{car94} find NGC 188 to be in a nearly circular orbit ($e = 0.07$) at a distance between 9.4 
to 10.8 kpc  from the Galactic center.
\citet{chu10} find the cluster to orbit between 6.97 and 9.03 kpc with a mean Galactocentric distance 
that is nearly the same as that of the Sun.  
They also find the orbit to have a low eccentricity of $e = 0.125$.
The differences between these two results are due to the Galactic models used and some updated 
observations used in the \citet{chu10} analysis.

For simplicity we place the cluster in a standard Galactic tidal field for the Solar neighborhood, 
having a circular orbit with a speed of 220 \kms~at a distance of 8.5 kpc from the Galactic center.
Adding the non-zero orbital eccentricity suggested by \citet{chu10} would introduce periods of
increased stellar escape rate due to the reduced tidal radius during perigalacticon, as well as 
the opposite effect at apogalacticon.  \citet{mad12} show that, after a Hubble time, 
a cluster in an elliptical orbit with an eccentricity of 0.5 and apogalacticon of 8 kpc will have 
$\sim$10\% less mass than the same cluster in a circular orbit at 8 kpc from the Galactic center.
Thus we assume that including the small eccentricity found by \citet{chu10} would result in only a 
minimal change to the mass of the model at 7 Gyr.
We also note that NGC 188 is currently out of the plane of our Galaxy, at $b=+22.$\degr$38$,
and its orbit crosses the galactic disk \citep{chu10}.  We do not account for any effects related to this disk crossing
(e.g., tidal disk shocks) in our model.
Disk shocks temporarily compress a cluster on a timescale shorter than the internal dynamical timescale, which accelerates all 
stars, some to velocities above the escape velocity (mostly those in the outer parts of the cluster).
\citet{ves97} show that for a cluster orbiting at 8.5 kpc, disk shocks produce only a minimal effect on the cluster mass.
This result is confirmed by \citet{kup10} who also show that the lifetime of a cluster on an 8.5 kpc orbit is much more strongly 
effected by a highly eccentric orbit than a highly inclined orbit.

\begin{figure*}[!ht]
\centering
\begin{tabular}{cc}
\includegraphics[width=0.4\linewidth]{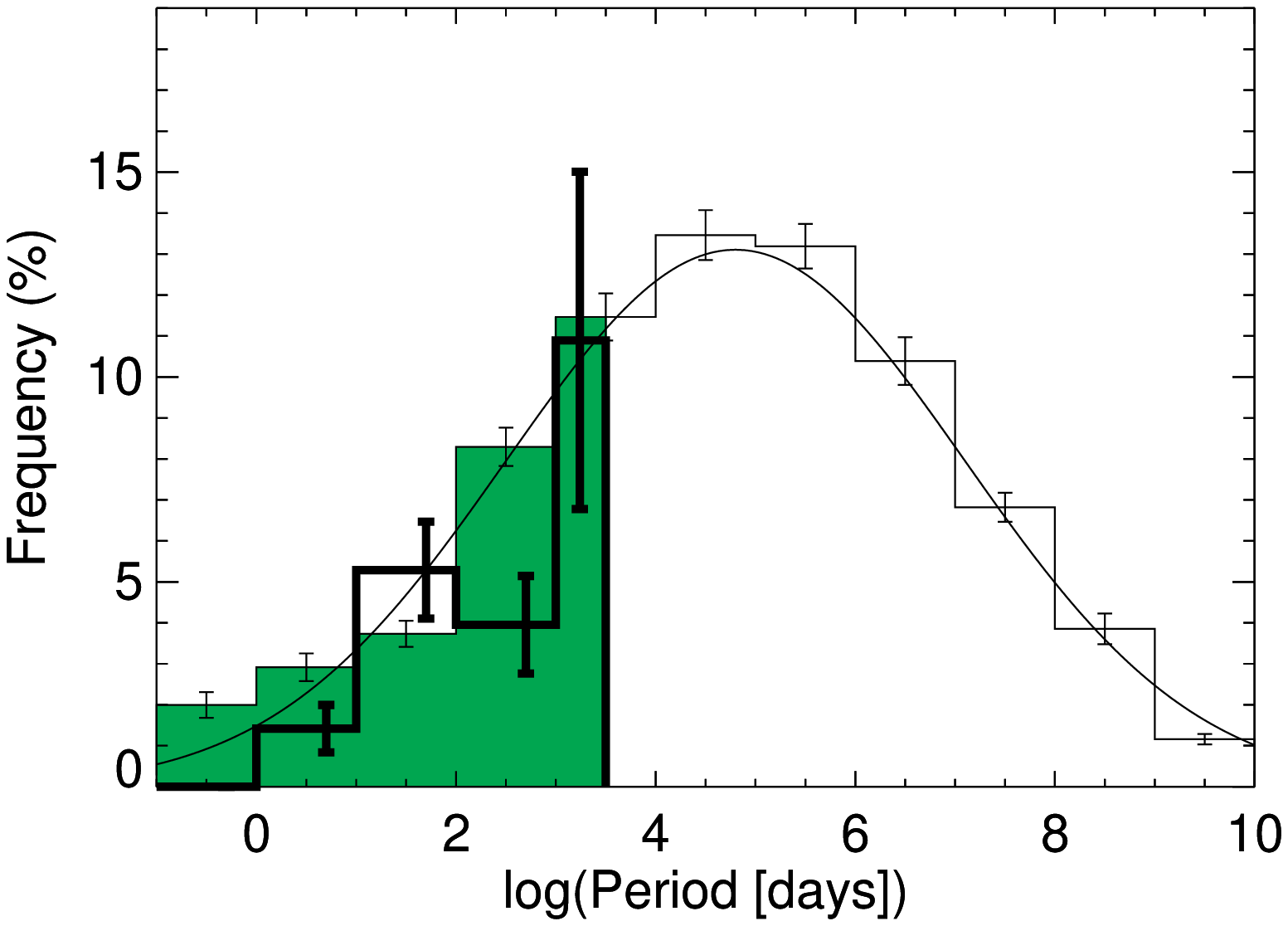} & \includegraphics[width=0.4\linewidth]{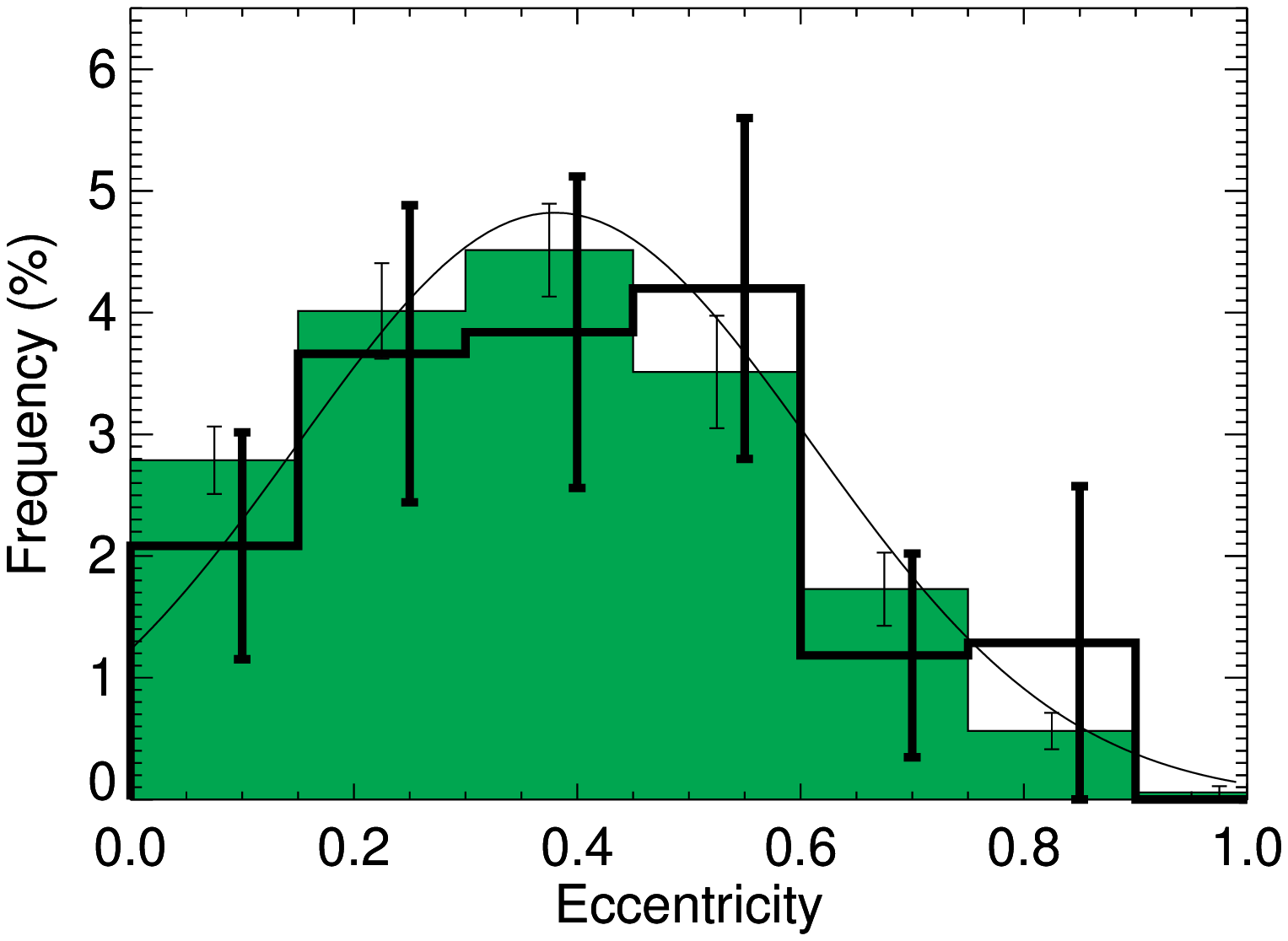} \\
\multicolumn{2}{c}{\includegraphics[width=0.85\linewidth]{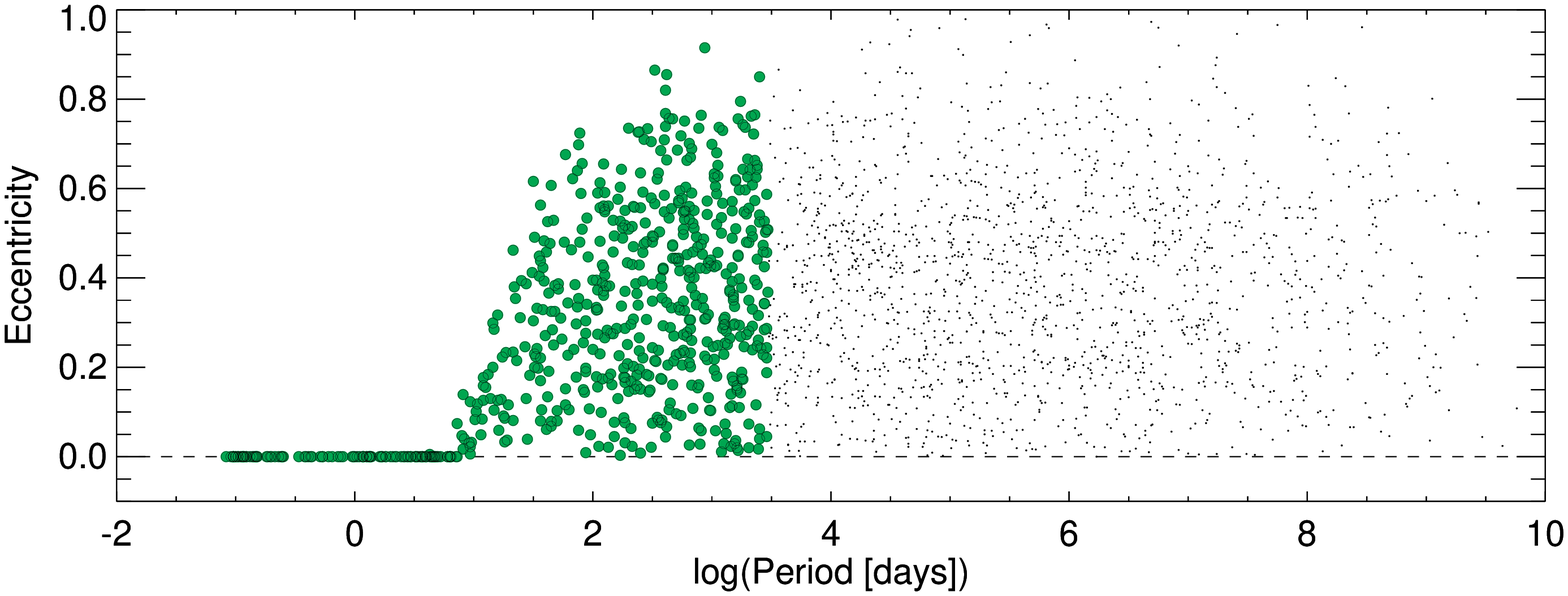}} \\
\includegraphics[width=0.4\linewidth]{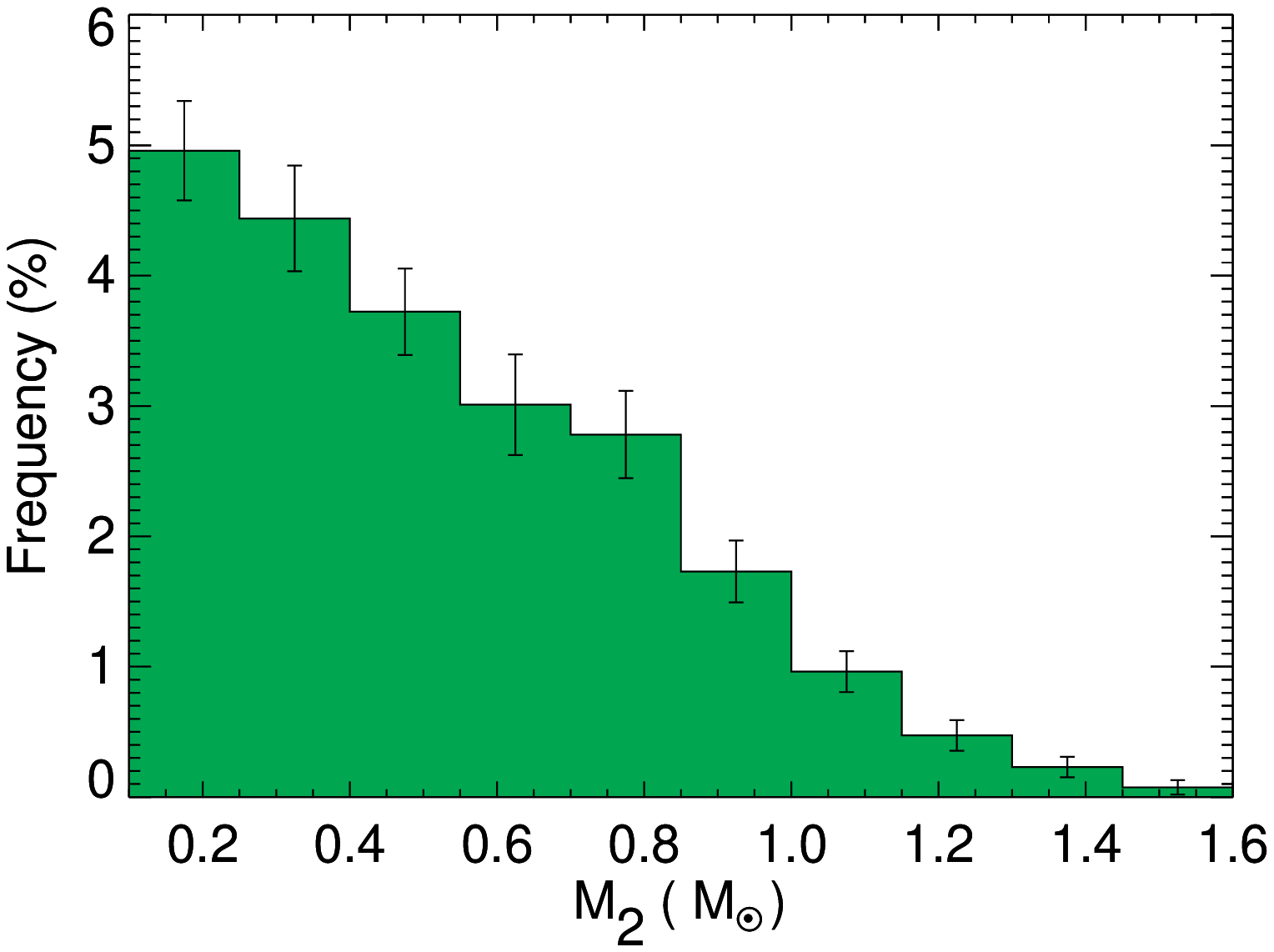} & \includegraphics[width=0.4\linewidth]{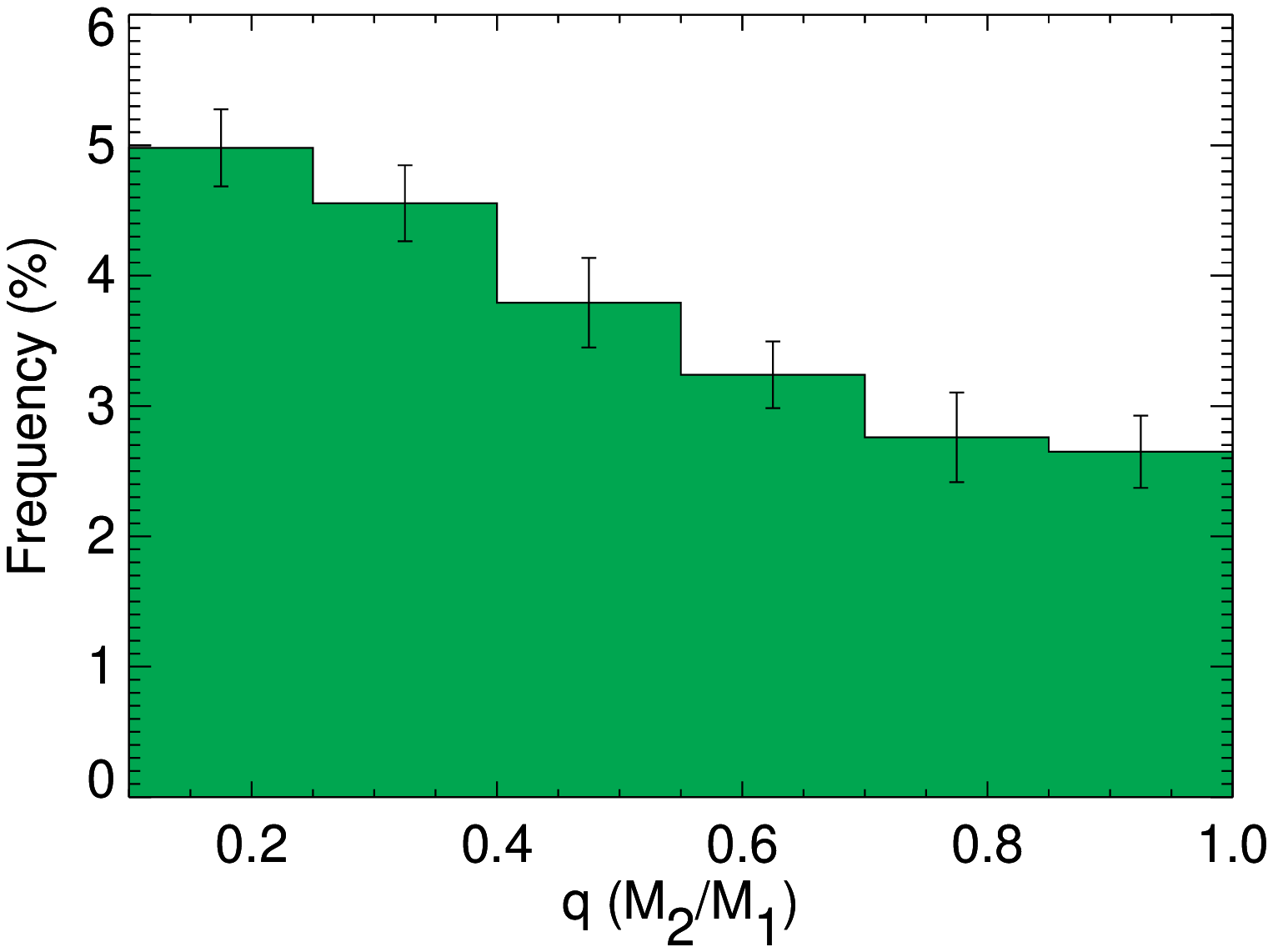} \\
\end{tabular}
\caption{\footnotesize
Initial distributions of solar-type binary orbital parameters for the NGC 188 model.  The 
distributions of period (top-left), eccentricity (top-right), 
secondary mass (bottom-left) and mass ratio (bottom-right) are shown in histogram form, 
and the eccentricity - log period ($e - \log (P)$) diagram is shown in the center.
Solar-type binaries within our observational limits (primary mass from 0.8 - 1.6 \Msolar~and $P < 3000$ days)
from the model are shown in the gray (green in the online version) filled histograms and points.
In the period and $e - \log(P)$ distribution plots we also show the 
full solar-type (0.8 - 1.6~\Msolar) sample in the thin-lined histogram and small points, respectively.
The observable sample in the eccentricity distribution plot excludes binaries with 
$P < P_{circ}$, where $P_{circ} = 10.2$~days 
\citep{mei05}.  
Each histogram bin shows the mean value for all simulations included in the model, and the 
error bars show one standard deviation above and below the mean.  The $e - \log (P)$ figure shows one
representative simulation.
The theoretical functions from which we draw the initial binary periods and eccentricities
in the model are shown in the black curves in the upper two plots (i.e., 
the log-normal period distribution from \citealt{duq91} and the Gaussian fit to the M35 eccentricity distribution).
We describe our method for choosing the initial mass-ratio 
distribution in Section~\ref{init_binary}.
Finally the observed M35 period and eccentricity distributions are plotted in thick-black histograms.
\label{initbinfig}
}
\end{figure*}

\subsection{Parameters of the Initial Binary Population} \label{init_binary}

The initial parameters defining a binary population in numerical simulations have long been derived from 
theoretical analyses and predictions, as we have lacked sufficiently detailed observations 
of young open cluster binary populations to define these parameters.
The WOCS survey of M35 \citep{gel10} now allows for a major step forward; we will use these
binaries to empirically guide our choices for the initial binary population in these simulations.

We don't yet know if all clusters, including NGC 188, form with similar binary populations.
However, as we will show in Section~\ref{final_binary}, an initial binary population based on 
observations of M35 will indeed match our observations of the NGC 188 binaries at 7 Gyr.
We discuss the implications of this result in Section~\ref{discuss}.

The M35 binaries in our sample show a rising distribution in period that agrees
with the log-normal distribution observed for solar-type binaries in the Galactic field \citep{duq91,rag10}. 
As we only detect binaries with periods $< 10^4$ days, we utilize these Galactic field 
binaries to extend the period distribution to longer periods.  Specifically we input a primordial 
binary population with a log-normal period distribution centered on log($P$ [days]) = 4.8
with $\sigma = 2.3$ (as defined by \citealt{duq91}).  We don't anticipate the initial soft binary
population to survive long against dynamical encounters, and so our results will be only minimally sensitive to this
extrapolation.

We observe M35 to have a MS binary frequency of 24\%~$\pm$~3\% out to periods of 10$^4$ days 
\citep{gel10}.  
We choose to set the initial binary frequency in the model to 27\% out to the same period cutoff. 
We allow the binaries to initially populate the entire period distribution, which corresponds 
to an initial total binary frequency of $\sim$70\% (16000 binaries).  However, the initially very long-period soft binaries 
($P\gtrsim10^7$ days) are quickly broken up through dynamical interactions.  
Thus the total binary frequency drops rapidly to 53\% within the first $\sim$30 Myr (see Figure~\ref{bfreqt}).  

The M35 binary eccentricity distribution 
(for binaries with $P_{circ} < P < 3000$ days, where $P_{circ} = 10.2\substack{+1.0 \\ -1.5}$ days from \citealt{mei05}) 
is consistent with a Gaussian distribution centered on $e = 0.38$ with $\sigma = 0.23$.
Thus we draw the initial binary eccentricities from this Gaussian function.
We note that this eccentricity distribution is again similar to that observed for solar-type binaries in the field.
xs
We have not yet performed an analysis of 
the observed secondary-mass or mass-ratio distributions for the M35 binaries.  
These distributions are generally the least certain observationally, at least for spectroscopic samples,
as the majority of observed spectroscopic binaries are single lined.
Observations of the 
Galactic field binary populations show the mass-ratio distribution to be either rising towards 
lower mass ratios \citep[e.g.][]{duq91} or uniform \citep[e.g.][]{maz03,rag10}. 
There is also evidence for a peak in the mass-ratio distribution at unity \citep{rag10,tok00}.
\citet{reg11} find a mass-ratio distribution in their binary sample that follows a
power law $dN/dq \propto q^\beta$, with $\beta = -0.50 \pm 0.29$, for binaries with primary masses 
between 0.25~\Msolar~and 6.5~\Msolar.

We choose to define the binary masses by first taking two masses, $M_1$ and $M_2$,
from the \citet{kro01} IMF, both more massive than $0.1$~\Msolar.  
This is approximately the hydrogen-burning limit, and no models below 0.1~\Msolar~were evolved
for the SSE code.
We then combine these masses ($M_{tot}$) and randomly 
choose a new mass ratio from a uniform distribution to define the primary and 
secondary masses for the binary, such that $M_1 + M_2 = M_{tot}$ and the secondary star has a mass
$M_2$ in the range $M_1 \leq  M_2 < 0.1$~\Msolar.  
The result of this procedure is shown in the bottom panels of Figure~\ref{initbinfig}.

This method does not produce the observed 
peak at mass ratios of unity for solar-type binaries; we will investigate here whether this peak 
can be produced through dynamical processes throughout the evolution of the cluster.
The solar-type 
binary mass-ratio distribution in our model agrees with the results of \citet{duq91}, who studied 
solar-type binaries in the field, and also with the results of \citet{reg11} and \citet{met09} who 
studied a wider range of binaries in different environments, all of whom find that the distribution of mass
ratios increases towards lower mass ratios.
(Although we note that \citealt{reg11} do not find a variation in the mass-ratio distribution
between different mass samples, which does result from our pairing method.)
We therefore proceed with this method for defining the component masses of the binaries in the NGC 188 model,
but we note that future models may benefit from a more thorough investigation of the effects of different 
methods for choosing binary component masses on the evolution of the binary population. 

The initial binary distributions are shown in Figure~\ref{initbinfig}.  
In this figure, as well as throughout
the paper we distinguish between the ``observable'' and ``total'' samples such that the observable 
sample includes only stars that are within the appropriate mass and period range for our radial-velocity 
observations of the true cluster(s), and total refers to the entire simulated cluster population.
Here the observable distributions are shown in filled histograms and larger points.

\subsubsection{Tidal Energy Dissipation Rate}

\begin{figure}[!t]
\plotone{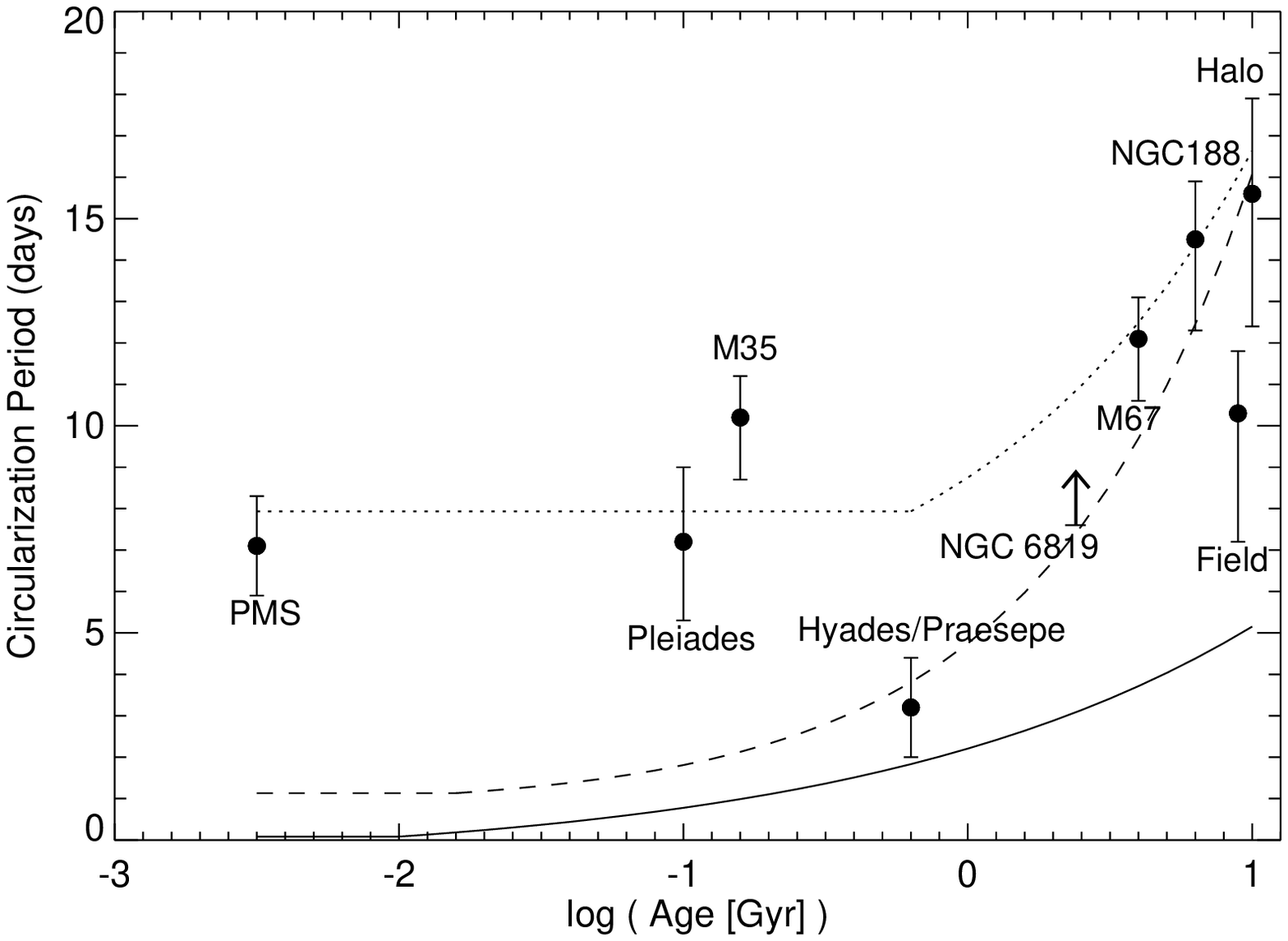}
\caption{\footnotesize
Circularization period as a function of age.
The labeled points reproduce the data from
\citet{mei05}, and we also add in the lower limit for the circularization period of NGC 6819 (2.4 Gyr) found by \citet{mil12}.
The lines show fits of exponential functions to the results from multiple population synthesis studies using the 
binary-star evolution \citep[BSE;][]{hur02} algorithm.
Specifically, the solid line shows the default BSE settings.  
The dashed line shows the result of increasing the convective damping term by a factor of 100.  
The dotted line shows the result of also adding the pre-main-sequence circularization algorithm from 
\citet{kro95b}. The default BSE settings result in circularization periods that are consistently less than 
observations of real binary population at any age.  With the addition of 
the pre-main-sequence circularization algorithm as well as the increased convective damping strength, the model
matches the observed circularization periods.
\label{psynthPcirc}
}
\epsscale{1.0}
\end{figure}

We digress briefly here to discuss the tidal treatment in the $N$-body code and the additional 
parameters that we impose on the model to match our observations.
The tidal circularization period is one of our best observational tools to study the effects of 
tides on a binary population.  \citet{mei05} define the circularization period as the period at which 
the best-fit circularization function (equation 1 in \citealt{mei05}) reaches an eccentricity of 0.01.
They use this method to study the circularization period of a number 
of binary populations of different ages, and we have reproduced the majority of their Figure~9 
here in Figure~\ref{psynthPcirc} with the labeled points.  They find a trend
of increasing circularization period with increasing age of the binary population, for ages $\gtrsim$1 Gyr.

In \citet{gel09b} we note that the tidal treatment in the $N$-body model produces circularization
periods that are well below those that are observed in real open clusters.  Indeed using the 
standard tidal energy dissipation rates in the $N$-body code results in a circularization period 
for solar-type binaries at 7 Gyr of $\sim$4 days, as compared to the observed value of 14.5$\substack{+1.4 \\ -2.2}$~days 
found by \cite{mei05} for similar binaries in NGC 188.  Therefore here we investigate methods for improving 
the agreement between the circularization periods derived from the $N$-body model and those from the observed 
binary populations through a series of population synthesis simulations using BSE \citep{hur02}.

The solid line in Figure~\ref{psynthPcirc} shows the circularization period as a function of 
time for a sample of isolated binaries evolved using the standard tidal prescription from the $N$-body and BSE codes. 
We derive all circularization periods from the population synthesis simulations using the same method as \citet{mei05},
and then fit exponential functions to these discreet circularization periods to produce the smooth curves shown here. 
The circularization periods from the standard BSE and $N$-body codes fall below those of nearly all observed clusters.
These low circularization periods suggest that the tidal energy dissipation rate may be 
underestimated within the model.  This has also been noted by \citet{bel08}, who 
suggest that the convective tidal damping rate should be increased by a factor of 50. 
We find that a factor of 100 increase is necessary in our model to be consistent with the lower limit on the tidal circularization 
period of NGC 188 at 7 Gyr (dashed line).

However for ages less than $\sim$4 Gyr the circularization period is still 
quite low as compared to observations of M35, the Pleiades and the pre-MS binary population.
Indeed, \citet{zah89} predict that pre-MS tidal circularization will result in
circularization periods between 7.2 and 8.5 days, in agreement with the empirical results of \citet{mei05}.

In order to achieve this early tidal circularization, we add the 
pre-MS binary evolution prescription suggested by \citet{kro95b}.\footnote{
In order to better fit the
``envelope'' in observed $e - \log(P)$ distributions of open clusters, we modify the $\lambda$ and $\chi$ values in the \citet{kro95b} algorithm
from 28 and 0.75, to 20 and 15, respectively.  
As described in \citet{kro95b}, $\lambda$ measures the length scale over which significant evolution of the orbital elements during 
the proto-stellar phase occurs, and $\chi$ measures the ``interaction strength'' between the two protostars in the binary 
system.
See \citet{gel12} for a comparison of the $e - \log(P)$ distribution in the \citet{hur05} model resulting from 
the $\lambda$ and $\chi$ values from \citet{kro95b} with the observed $e - \log(P)$ distribution of NGC 188.}
The results of this combination of both the increased convective tidal damping coefficient and the \citet{kro95b} pre-MS 
binary evolution prescription is shown in the dotted line.
(For comparison, using the pre-MS binary evolution prescription without increasing 
the convective tidal strength results in essentially the same $P_{circ}\sim8$~days at all times.)

We note that this procedure is simply the \textit{ad hoc} addition of two mechanisms, and here we 
use them together only to (a) achieve the desired initial form to the $e - \log(P)$ distribution 
and (b) maintain a circularization period that is consistent with the observations of \citet{mei05} out to the age of NGC 188.
Future work is necessary to investigate whether such a ``hybrid'' style model can be developed into 
a more comprehensive tidal theory (as has also been suggested by \citealt{mat92} and discussed by \citealt{mei05}).

In summary, to initialize the binaries in our NGC 188 model, we first choose orbital parameters independently from the 
observationally defined distributions discussed above.  In rare cases we modify these initial orbital parameters to 
avoid unphysical systems, (e.g. those that would collide or merge in a single orbit).
We then modify the binaries according to the pre-MS binary evolution of 
\citet{kro95b}.  The resulting distributions of orbital parameters are shown in Figure~\ref{initbinfig}.
As the binaries evolve within the model they are subject to the increased tidal damping factor 
described above.

\section{Color-Magnitude Diagram, Mass and Structure of the NGC 188 Model and Observations at 7 Gyr} \label{SOBS}

\begin{figure*}[!t]
\plotone{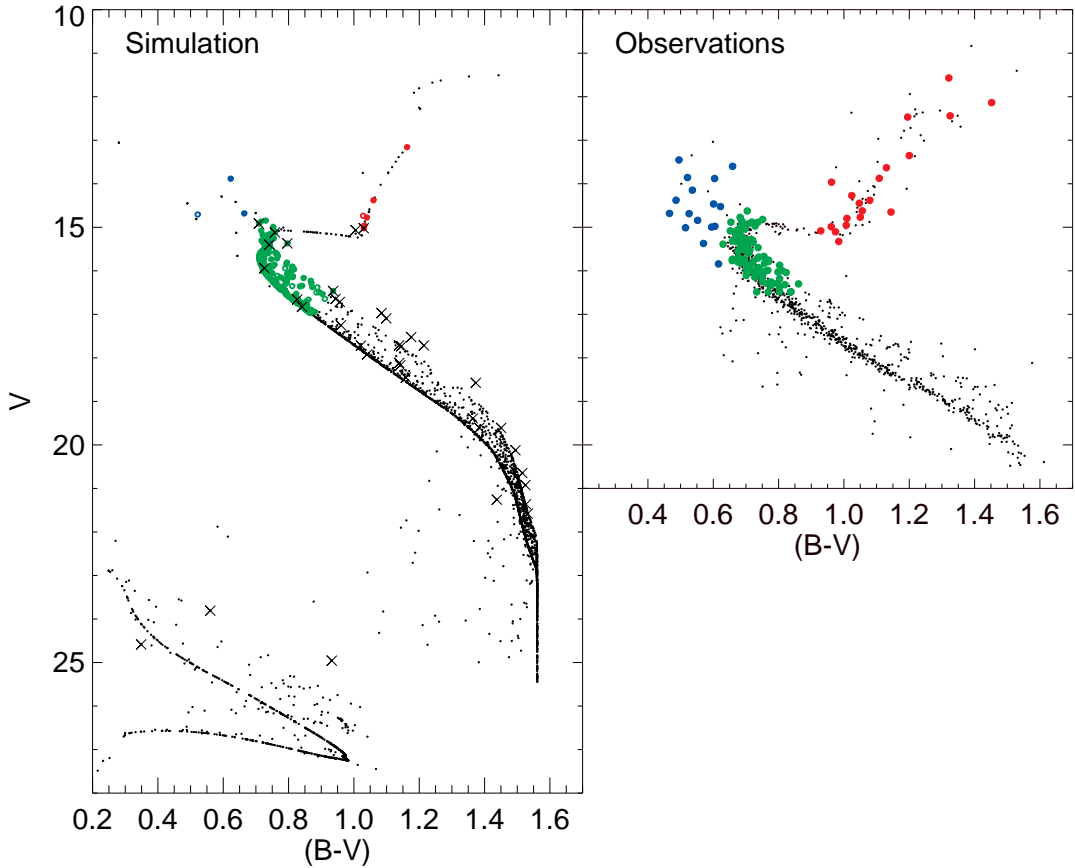}
\epsscale{1.0}
\caption{\footnotesize
Color-magnitude diagrams (CMDs) of a representative simulation from the NGC 188 model (left) and observations of the true cluster (right).
For both panels, the detectable binaries ($V < 16.5$ and $P < 10^4$ days) are plotted in the larger filled 
circles with the main-sequence binaries in light gray, giant binaries in dark gray and blue straggler 
binaries in black (green, red and blue in the online version).
Solar-type binaries with periods longer than $10^4$~days in the simulation are plotted in open circles.
Triples are plotted in crosses. All other stars are plotted in black points.  
In the observed CMD, we show all cluster members from the \citet{gel08} survey and extend the sample to 
fainter magnitudes using the proper-motion members ($P_{PM} > 50$\%) from \citet{pla03}.
The proper motions extend to a faint limit of $V=21$, but begin to become incomplete at $V \sim 19$.  
For the simulation, binaries and triples are plotted as single points, 
showing their combined magnitudes and colors.  The distance modulus of $(m-M)_V = 11.44$
and reddening of $E(B-V) = 0.09$ of \citet{sar99} are employed to place the simulation at the same 
distance and color as NGC 188. 
\label{CMD}
}
\end{figure*}

We evolve the stellar population, as defined above, dynamically with the \texttt{NBODY6} code to an age of 7 Gyr to create a simulation of NGC 188.
As discussed above, we ran twenty simulations each with a different realization of the same initial parameter distributions,
and for much of the analysis that we discuss here and in the following sections, we combine the results of these twenty simulations
to form the NGC 188 model.

First we discuss the 7 Gyr CMD, Figure~\ref{CMD}. 
In the left panel of this figure, we show one representative simulation;  the other 19 simulations
have qualitatively the same features, although the specific numbers of stars in various populations 
(e.g. BSs and triples) vary.  Here we briefly compare this simulation to the observed CMD for NGC 188, shown in the right panel of same figure.

The magnitudes for each star in a given simulation are calculated in the same manner as in \citet{hur05},
using the bolometric corrections of \citet{kur92} for MS and giant stars and of \citet{ber95} for WDs.
We use the distance modulus $(m-M)_V$ = 11.44 and reddening $E(B-V)$ = 0.09 found by 
\citet{sar99} to place the simulation at the appropriate distance and color.  
In both CMDs, we show all stars with the small black points and highlight 
binaries within our observational detection limits with larger 
points, using light gray for MS binaries, dark gray for red giant branch (RGB) binaries, and black for BS binaries
(green, red and blue in the online version).
We use this same gray-scale (color) coding in our analyses of these respective binary populations throughout the paper.

At the distance of NGC 188, the components of a circular binary at the hard-soft boundary 
($P \sim 10^{6.5}$ days for solar-type stars at 7 Gyr) would be separated by $\lesssim0.01$ arcseconds on the sky.  
Therefore, no binaries (or higher order systems) are resolvable from ground-based 
observations, and here we treat all binaries (and higher order systems) in the model as unresolved objects.

The model qualitatively reproduces the MS single and binary sequences quite closely
as compared to the observed CMD.  We see a width to 
the MS in both CMDs as a result of the rich binary population of the cluster.  
(We will discuss the binaries in detail in Section~\ref{final_binary}.)  The 
cluster turnoff is at $V \sim 14.8$ in both CMDs, which corresponds to a mass of $\sim$1.1~\Msolar.  
The MS luminosity function from the model is also consistent with that of the proper-motion 
members from \citet{pla03} in both number and form (for $V \lesssim 19$, below which the \citealt{pla03} 
catalog becomes incomplete).

However, there are 
two obvious discrepancies between the observed and simulated clusters.  First the model
does not predict the large long-known scatter on the RGB observed in NGC 188 
\citep[e.g.,][]{mcc77,twa78,gel08}.  The source of this observed scatter is not due to 
the precision of the photometry.  The $BV$ photometry shown here is precise to 0.01 mag \citep{pla03}.
We have not added photometric errors to the simulated CMD.
The origin of this observed scatter on the NGC 188 RGB, also seen in other clusters 
(e.g., NGC 6819, \citealt{hol09}), remains unknown.

Second, the model clearly has fewer BSs at 7 Gyr than the true cluster.  We will discuss the BSs in 
detail in Section~\ref{final_BS}.

\begin{figure}[!t]
\plotone{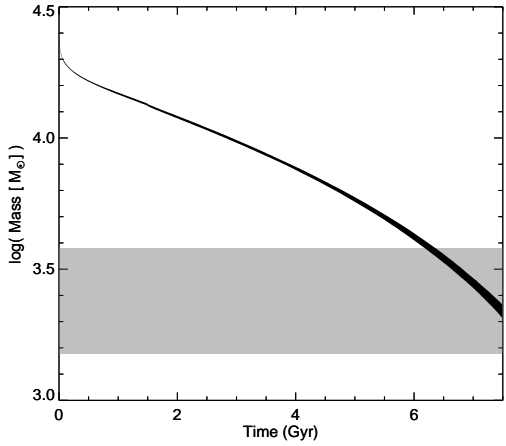}
\epsscale{1.0}
\caption{\footnotesize
Logarithm of the mean total cluster mass as a function of time in the NGC 188 model. 
The black line is centered on the mean total mass at a given time from our twenty simulations, and the width of 
the line shows one standard deviation above and below the mean.
The light-gray horizontal band shows the range in mass derived empirically for NGC 188 of 1500~\Msolar~to~3800~\Msolar~\citep{bon05,gel08,chu10}.
The model begins with a mean total mass of 23610~\Msolar~(with a standard deviation of 40~\Msolar~between the simulations),
and at 7 Gyr the mean total mass is reduced to 2850~\Msolar~(with a standard deviation of 120~\Msolar) through stellar evolution and evaporation processes.
The mean mass of the model at the age of NGC 188 agrees with the observed mass estimates.
\label{massvtime}
}
\end{figure}

Next, in Figure~\ref{massvtime} we show the mean total mass of the model as a 
function of time.  The mean initial mass of all twenty simulations is 23610~\Msolar, with a standard deviation of 40~\Msolar.
At 7 Gyr the simulated clusters have lost on average $\sim$88\% of this mass through evaporation processes and stellar evolution, 
leaving a mean final mass of 2850~\Msolar, with a standard deviation of 120~\Msolar. 
This value is in good agreement with empirical mass estimates for NGC 188 that range
from 1500~\Msolar~to~3800~\Msolar~\citep{bon05,gel08,chu10}.

Next we investigate the cluster structure at 7 Gyr.  For this and later analyses we 
analyze each simulation in the NGC 188 model as it might appear to an observer on Earth.
NGC 188 has Galactic coordinates of $l=122.$\degr$85$ and $b=+22.$\degr$38$ \citep{pla03}.  
Each simulation begins with the X-direction facing away from the Galactic center and the cluster orbits in the X-Y plane.
We rotate each simulation according to the Galactic coordinates such that the transformed $Y'- Z'$ plane becomes the plane of the sky, 
and the $X'$ direction is the line-of sight.  
We then use this observed line-of-sight to NGC 188 in order to investigate the cluster 
structure at 7 Gyr, and for subsequent analyses of the model where appropriate.

\begin{figure}[!t]
\plotone{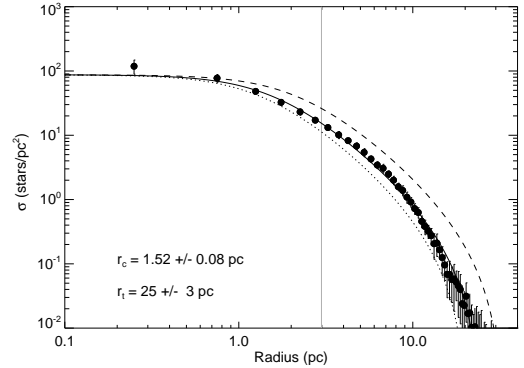}
\caption{\footnotesize
Stellar surface density as a function of radius from the cluster center at 7 Gyr in the model, projected into
the true line-of-sight towards NGC 188. Each point shows the mean value for all simulations within 
the given radial bin, and the error bars show one standard deviation above and below the mean.
We include only stars (and binaries) from the simulation with $V < 21$, to match the limiting
magnitude of the \citet{pla03} proper-motion study of the true cluster.
The solid black line shows a standard  King model fit to these data, which yields a core radius 
$r_c = 1.52 \pm 0.08$~pc and a tidal radius of $r_t = 25 \pm 3$~pc.  
For comparison we also show two 
King model fits from the literature to the observed density profile of NGC 188. 
The King model fit of \citet{bon05} is shown in the dotted line, who find NGC 188 to have a $r_c = 1.3$~pc and 
$r_t = 21$~pc, and the King model fit of \citet{chu10} is shown with the dashed line, who find $r_c = 2.1$~pc
and $r_t = 34.4$~pc (both normalized to the same central density as our fit to the NGC 188 model).   
The vertical line shows the mean half-mass radius of the model at 7 Gyr. 
\label{Ndensity}
}
\epsscale{1.0}
\end{figure}

In Figure~\ref{Ndensity} we show the stellar surface density as a function of radius from the 
cluster center at the age of 7 Gyr.  Binaries and higher-order systems are treated as individual objects.
Each point in this figure shows the mean stellar density at the given radius, averaged over all simulations.  
The error bars show the standard deviations of these values.
This profile is fit with the empirical three-parameter \citet{kin62} model, 
as shown by the solid black line ($\chi^2_{red} = 1.9$).  The fit results in a core radius of 
$1.52 \pm 0.08$~pc
and a tidal radius of 
$25 \pm 3$~pc.

With the dotted and dashed black lines in Figure~\ref{Ndensity}, we show two recent results analyzing the 
observed stellar density profile in NGC 188.  The dotted line shows the results of 
\citet{bon05} who used 2MASS data to derive a core radius of $1.3\pm1$~pc and a tidal radius of 
$21\pm4$~pc.  The dashed line shows the results of \citet{chu10} who use the NGC 188 proper-motion 
members from \citet{pla03} to derive a core radius of $2.1\substack{+0.9 \\ -0.6}$~pc and a tidal 
radius of $34.4\substack{+16 \\ -10}$~pc.  The values for the NGC 188 
$N$-body model fall comfortably between the results of these two studies.

\begin{figure*}[!t]
\epsscale{1.0}
\plotone{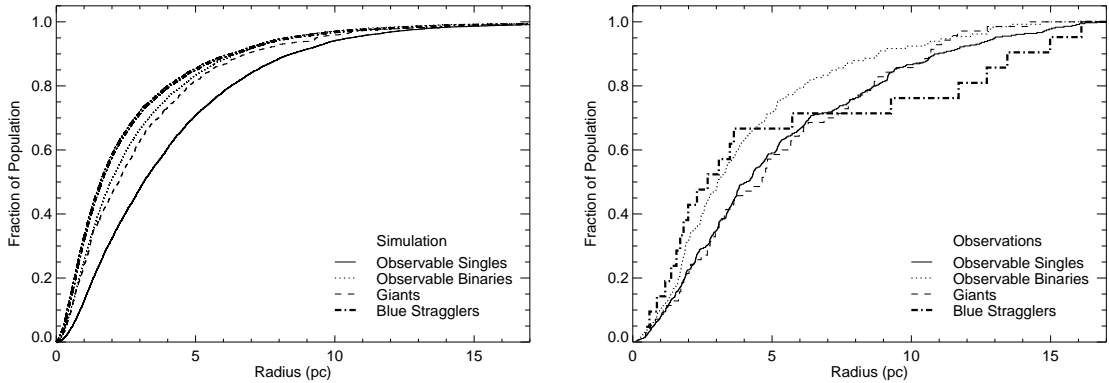}
\caption{\footnotesize
Radial cumulative distributions of cluster populations from the NGC 188 model (left; in observed line-of-sight projection) 
and observations (right; reproduced from \citealt{gel08}).
The cluster is divided into the observable singles (solid line), observable binaries (dotted line), giants (dashed line) and
blue stragglers (dot-dashed line).
The solar-type binaries are centrally concentrated with respect to the solar-type single stars in the model and observations, due to 
two-body relaxation processes.
The simulated blue stragglers do not show the same bimodal spatial distribution that is observed in NGC 188.
\label{cumR188}
}
\epsscale{1.0}
\end{figure*}

In Figure~\ref{cumR188} we compare the 
cumulative radial distributions of the observable single, binary, RGB and BS populations in the model with
those of the true cluster from \citet{gel08}.  We include the BS radial profile here for completeness, but we will wait to
discuss the BS population until Section~\ref{final_BS}.
For Figure~\ref{cumR188} and all subsequent cumulative distribution plots presented in this paper, we use the union of 
all twenty simulations to construct the cumulative distribution functions for the NGC 188 model.
The solar-type binaries in the model are centrally concentrated with respect to the 
solar-type single stars at the $>$99\% confidence level, in agreement with the true cluster.
This result also holds for any individual simulation.
The solar-type binaries are on average more massive than the single stars, and this result reflects the fact 
that the model is mass segregated at 7 Gyr, as has also been observed in the true cluster \citep{sar99,kaf03,gel08}.

The RGB stars in the model appear to be more centrally concentrated than the single stars when combining all twenty simulations (as shown in Figure~\ref{cumR188}).
This is in contrast to the observed radial distribution of the RGB stars in NGC 188, which follow a nearly identical radial distribution to 
the single stars in the true cluster.  
However for any given simulation, a K-S test does not return a significant distinction between the radial distributions
of the RGB and single stars (given the relatively small number of RGB stars per simulation).

\section{The 7 Gyr Main-Sequence and Giant Binary Populations} \label{final_binary}

\subsection{Binary Frequency}

\begin{figure}[!t]
\centering
\plotone{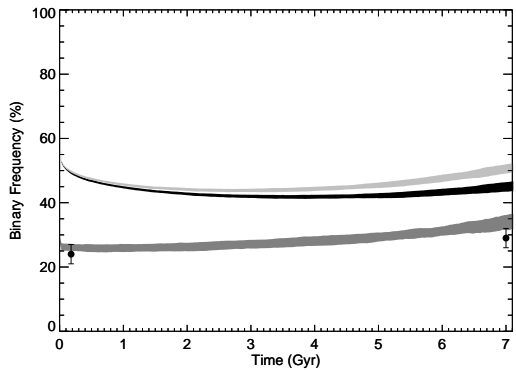}
\caption{\footnotesize
Binary frequency as a function of time in the NGC 188 model.  The total binary frequency is shown with the black line, 
the total main-sequence binary frequency with the light-gray line, and the observable main-sequence binary frequency
with the dark-gray line (M$ > 0.9$~\Msolar~and $P < 10^4$~days).
All lines are centered on the mean values of all twenty simulations, and the widths are one standard deviation above and 
below each mean.
Also shown (as points) are the observed main-sequence hard-binary 
frequencies of M35 (24\%~$\pm$~3\% at 180 Myr) and NGC 188 (29\%~$\pm$~3\% at 7 Gyr), both for $P<10^4$~day binaries.  
The observable main-sequence binary frequency in the  model is consistent with those of both clusters at their respective ages.  
We also note that the total binary frequency drops below the main-sequence binary frequency as the cluster evolves due to the population of binaries with white-dwarf primaries, 
which has a low binary frequency. (In detached binaries, mass-loss on the giant branch increases the orbital separation and thereby increases 
the probability for dynamical encounters resulting in exchanges or disruption.)
\label{bfreqt}
}
\epsscale{1.0}
\end{figure}

At 7 Gyr in the twenty simulations the mean observable MS hard-binary ($P < 10^4$ days) frequency is 
33.5\%, with a standard deviation of 2.8\%.
This result is in agreement with the observed MS hard-binary frequency of NGC 188 of 29\%~$\pm$~3\%
\citep[within the same period and mass range;][]{gel12}.
As is clear from Figure~\ref{bfreqt} the model predicts that the binary frequency will change only slightly
over 7 Gyr of evolution.  Moreover, a cluster with a binary frequency consistent with that of M35 at 180 Myr will 
evolve to have a binary frequency consistent with NGC 188 at 7 Gyr.

The mean hard-binary frequency of the RGB stars in the model is 31\% with a standard deviation of 6\%,
indistinguishable from that of the model MS stars.  This binary frequency also agrees with that of the true cluster;
\citet{gel12} find the NGC 188 giants to have a hard-binary frequency of 34\%~$\pm$~9\%.

\begin{figure}[!t]
\centering
\plotone{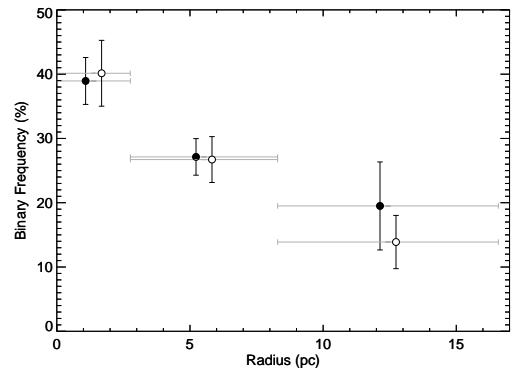}
\caption{\footnotesize
Binary frequency as a function of radius at 7 Gyr for the NGC 188 model and observations.
For both the model and observations, the sample includes main-sequence and giant stars combined.
Filled circles show the model binary frequencies and open circles show the observed binary frequencies.
Errors bars on the observations
show the Poisson counting uncertainties.  The points from the model show the mean values from all simulations, 
and the error bars show one standard deviation above and below the means, respectively.
The horizontal bars indicate the range
in radius from the cluster center for which the binary frequency is calculated.  
(The bins used for the observed and simulated data are identical; the plotted points are shifted along the x-axis for clarity.)
The binary frequency increases towards the core in the model due to two-body relaxation processes.
Remarkably, the binary frequency in the model matches that of the true cluster at each bin in radius.
\label{bfreqr}
}
\epsscale{1.0}
\end{figure}

Again we see clear evidence of mass segregation (in both the simulated and true cluster) 
when we examine the 
hard-binary frequency of the observable MS and RGB populations combined as a function of radius from the cluster center (Figure~\ref{bfreqr}).
We exclude the BSs from the analysis, because the radial dependence of their binary frequency 
may be tied to their formation mechanisms.
Here the model is shown in the filled circles and is compared to the result from \citet{gel12} for the true cluster 
MS and RGB stars, shown in open circles.  
The agreement is remarkable.

\vspace{1em}
\subsection{Distributions of Periods, Eccentricities, Secondary Masses and Mass Ratios} \label{final_binary.dist}

\begin{figure}[!t]
\plotone{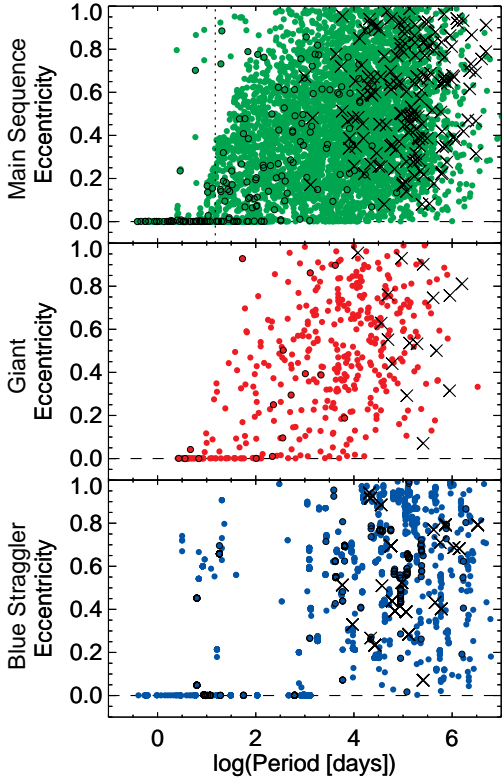}
\caption{\footnotesize
Eccentricity plotted against the logarithm of the period for the main-sequence (top), giant (middle) 
and blue straggler (bottom) binaries in the model. 
For the main-sequence and giant binaries, we show all binaries present in all simulations at 7 Gyr.
The integrated sample of blue stragglers shown in the bottom panel
contains all blue-straggler binaries present in the model at each snapshot interval between 6 and 7.5 Gyr. 
The circularization period $P_{circ}=14.5$ days \citep{mei05}
is plotted for reference with the dashed vertical line in the main-sequence panel.
In all three panels, we circle inner binaries of triple systems,
and plot the outer orbital parameters for triples with crosses.
\label{elogPfig}
}
\epsscale{1.0}
\end{figure}

\begin{figure}[!t]
\plotone{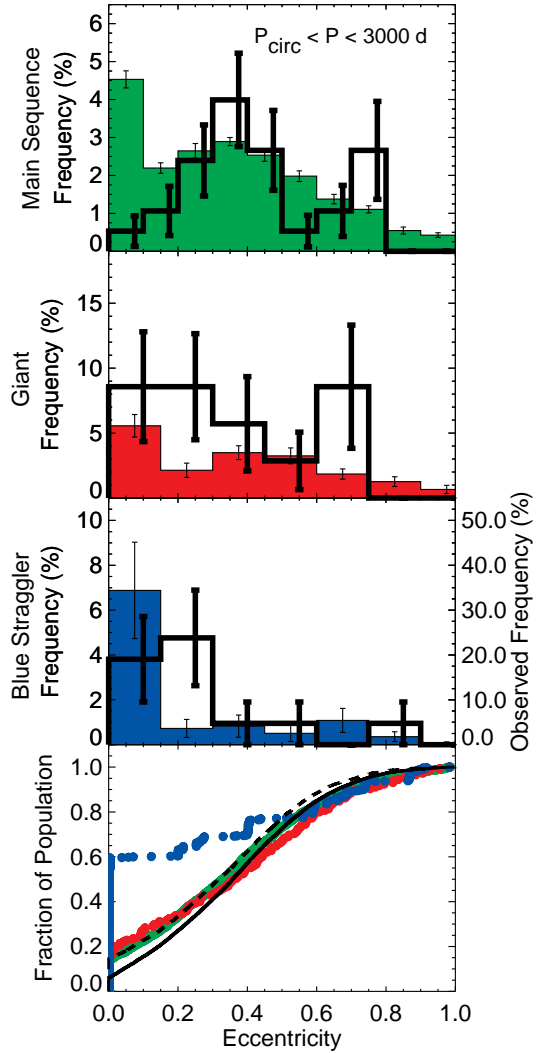}
\caption{\footnotesize
Eccentricity distributions for the main-sequence, giant and blue straggler binaries with $P_{circ}< P < 3000$ days
(where $P_{circ} = 14.5$ days; \citealt{mei05}). 
The period limit is set empirically to exclude binaries whose orbits have been circularized by tides, and to extend to 
our completeness limit in orbital solutions for the NGC 188 sample. 
In the top three panels the model is shown in filled histograms, showing all solar-type main-sequence and giant binaries 
present at 7 Gyr and the integrated blue straggler sample.
Each histogram bin represents the mean number of binaries in all twenty simulations 
within the given eccentricity range, and the error bars show the standard errors of these means.
For comparison we plot the observed NGC 188 distributions in the thick-lined histograms, respectively.
The observed NGC 188 blue straggler frequency distribution is divided by 5 for clarity, and the true observed frequency is shown on the 
right y-axis.
In the bottom panel we plot the cumulative distributions for the main-sequence,
giant and blue straggler samples in the model in light gray, dark gray and black (green, red and blue in the online version).
For comparison we also plot the solar-type main-sequence eccentricity distribution of the initial population (thin solid line) and 
that resulting from evolving the same initial binaries in 
isolation for 7 Gyr (dashed line).
\label{efig}
}
\epsscale{1.0}
\end{figure}

\begin{figure}[!t]
\plotone{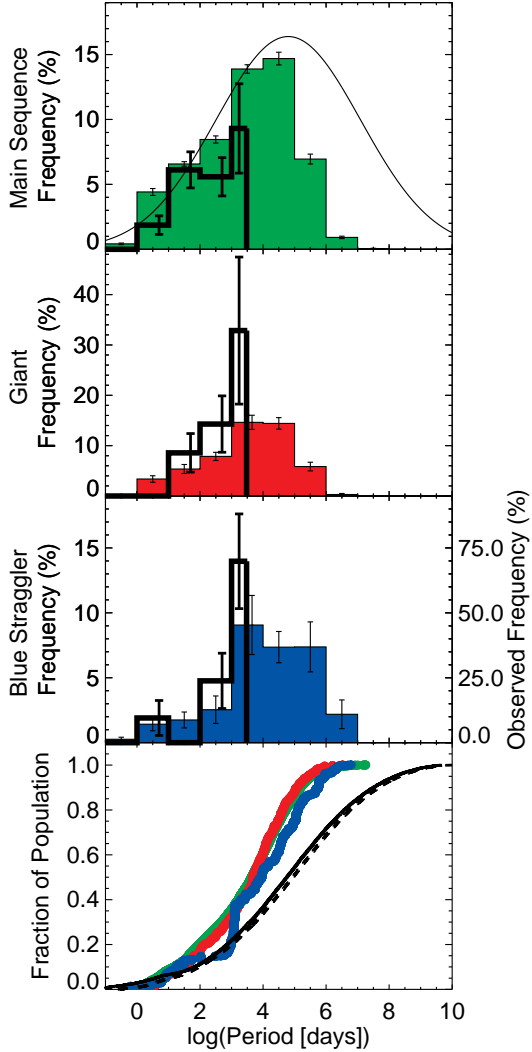}
\caption{\footnotesize
Period distributions for the main-sequence, giant and blue straggler binaries.
The plots are of the same format as Figure~\ref{efig}, except here we do not limit
the model by period.  (The observations extend to a period of 3000 days.)
Additionally we show the full log-normal period distribution from which we chose the initial 
binary periods in the top panel, normalized to the 7 Gyr solar-type main-sequence binary frequency.
\label{Pfig}
}
\epsscale{1.0}
\end{figure}

\begin{figure}[!t]
\plotone{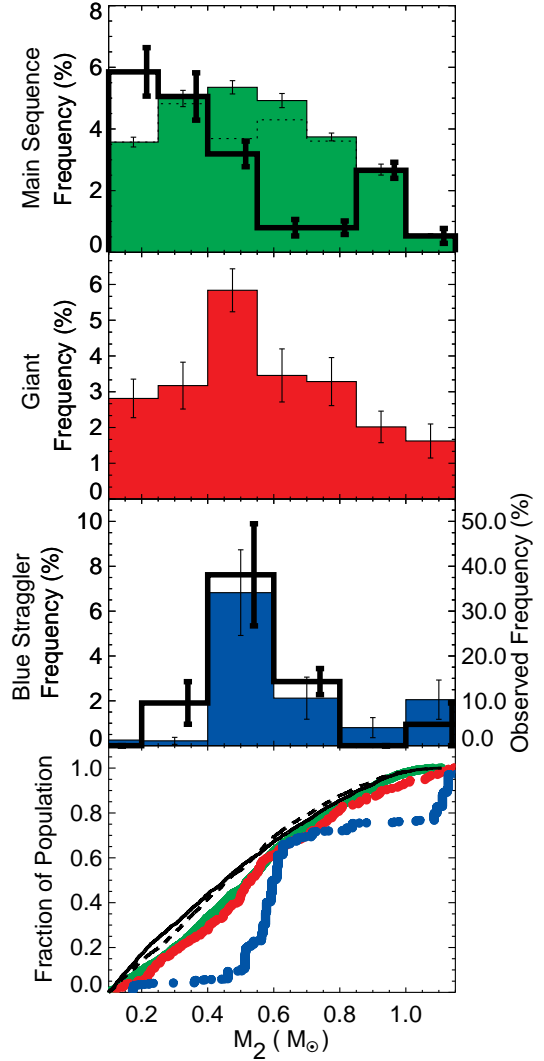}
\caption{\footnotesize
\label{M2fig}
Secondary-mass distributions for the main-sequence, giant and blue straggler binaries.
The plots are of the same format as Figure~\ref{efig}, except here we include
all solar-type observable binaries with $P < 3000$ days. Additionally 
in the main-sequence panel we show the distribution excluding the unphysical MS-WD binaries with 
the dotted line. Note the plots begin at 
0.1~\Msolar, as this is the minimum initial stellar mass evolved within the $N$-body code.
(\cite{gel12} do not derive a secondary-mass distribution for the giants.)
}
\epsscale{1.0}
\end{figure}

\begin{figure}[!t]
\plotone{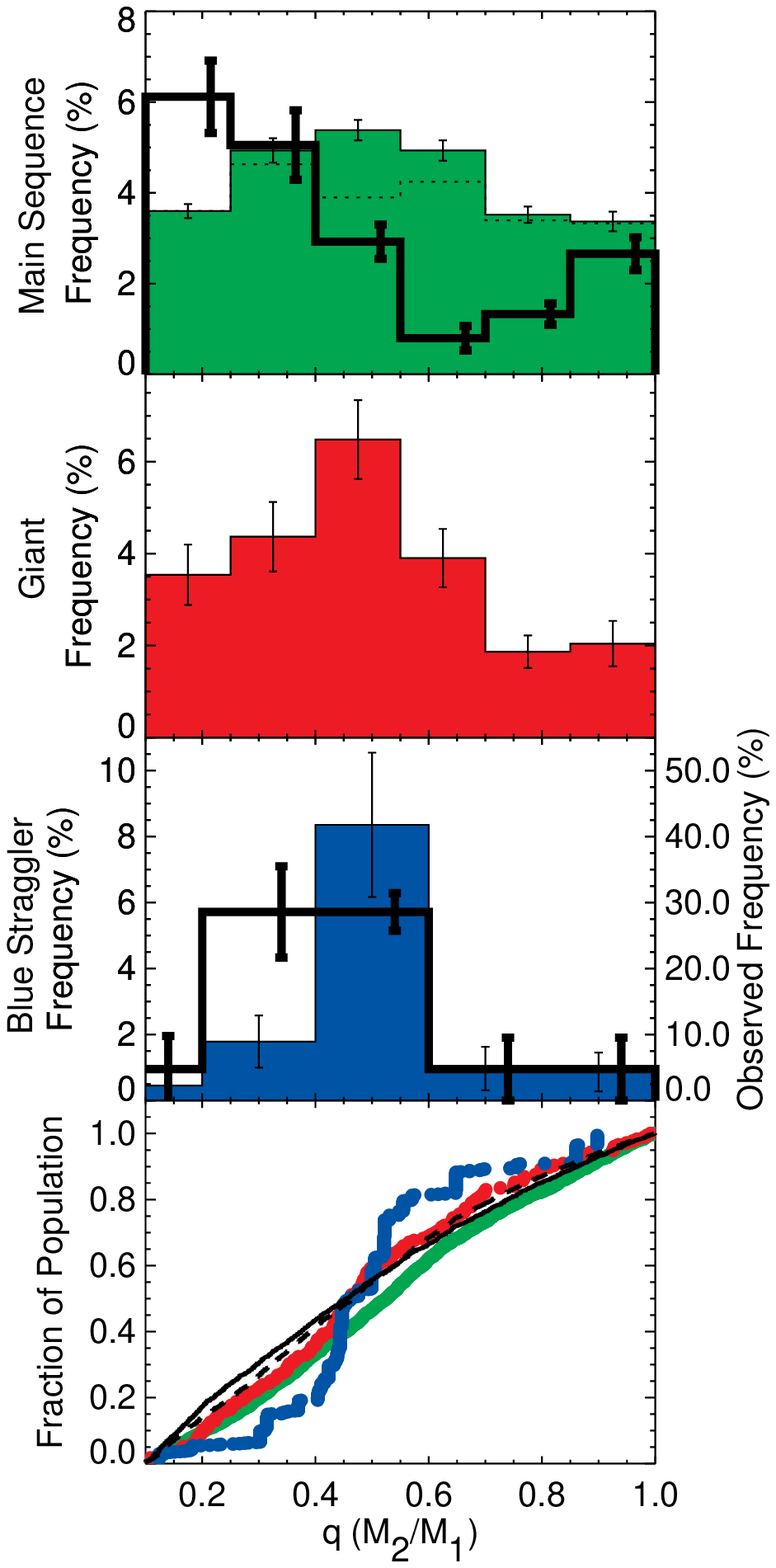}
\caption{\footnotesize
\label{qfig}
Mass-ratio distributions for the main-sequence, giant and blue straggler binaries.
The plots are of the same format as Figure~\ref{M2fig}.
}
\epsscale{1.0}
\end{figure}

In Figures~\ref{elogPfig}~-~\ref{qfig} we show the eccentricity - log period ($e - \log(P)$) distribution and the distributions of 
eccentricities, periods, secondary masses and mass ratios
for the observable solar-type binaries in the model as compared to observations of similar binaries in NGC 188.
In all figures, the MS, RGB and BS binaries from the model are plotted in light gray, dark gray and black (green, red and blue in the online version).
For the selection of the RGB and BS binary populations, only one star must be a 
RGB or BS star, respectively.
For completeness, we include the BSs in Figures~\ref{elogPfig}~-~\ref{qfig}, 
but we wait to discuss the BS population and the selection of the BS sample until Section~\ref{final_BS}.
Here we focus on the MS and RGB binaries.

In Figure~\ref{elogPfig} we show the $e - \log(P)$ diagram for the observable 
MS (top), RGB (middle) and BS (bottom) stellar populations in the model.  The MS and RGB $e - \log(P)$ distributions
are in close agreement with the observations (Figure 3 in \citealt{gel12}).
The mean 7 Gyr circularization period over all of our simulations is $17.3 \pm 2.1$~days, consistent with the value of
14.5$\substack{+1.4 \\ -2.2}$~days found by \cite{mei05} for NGC 188.

However, we notice a significant population of circular MS binaries with $P > P_{circ}$ in the model, which is 
also seen in the eccentricity distribution shown in Figure~\ref{efig}. 
Such binaries are not observed in NGC 188 \citep{gel12},  the Galactic field \citep{duq91,rag10}, or 
any of the open cluster binary populations studied by \citet{mei05}.  We will return to these binaries in Section~\ref{discuss_lPcirc}.
Here we simply note that on average in a given simulation $<$1 of these long-period circular binaries had an initial 
eccentricity $<0.01$.
The vast majority started with non-zero eccentricities, but were circularized 
during a CE episode which left a WD secondary star.
If we do not include these post-CE systems, the MS eccentricity distribution of the model is statistically indistinguishable from 
the observed distribution.

At 7 Gyr the solar-type MS binary period distribution in the model is also consistent with that observed in NGC 188.
The simulated MS period distribution in Figure~\ref{Pfig} rises towards our observational
detection limit as does the observed distribution. 
Including only binaries that are within our observational completeness limit of $P < 3000$ days,
the observed and simulated period distributions are statistically indistinguishable.

The MS secondary-mass and mass-ratio distributions in the model (Figures~\ref{M2fig}~and~\ref{qfig}) have forms somewhat different from
those of the true cluster. Both distributions in the true cluster rise towards lower-mass companions
(although \citealt{gel12} note that they cannot formally distinguish the observed mass-function distribution, the observable quantity, from 
a distribution derived by choosing binary component masses from a uniform mass-ratio distribution).
In both the MS secondary-mass and mass-ratio distributions from the model, we see distributions that rise until 
$M_2\sim$0.5~\Msolar~and  $q\sim$0.5, followed by a turnover towards lower masses and mass ratios.
The peak in both distributions is in part a result of WD companions, the vast majority of which are members of the unphysical 
long-period circular MS-WD binaries discussed briefly above.
If we remove these MS-WD binaries (dotted lines in the MS panels of Figures~\ref{M2fig}~and~\ref{qfig}), the distributions in 
secondary-mass and mass-ratio are both consistent with a uniform distribution, respectively.

We see a similar peak at about 0.5~\Msolar~in the secondary-mass and mass-ratio distributions for the RGB binaries.  This peak
is also primarily due to WDs.  Furthermore $\sim$60\% of these RGB - WD binaries evolved from post-CE MS - WD binaries with long-periods and 
circular orbits, similar to those discussed above.  Nearly all of the remaining 40\% are also evolved post-CE MS -WD binaries, but with 
$P < P_{circ}$. Again, we will discuss these long-period circular post-CE MS-WD binaries in detail in Section~\ref{discuss_lPcirc}

The observed NGC 188 MS mass-ratio distribution shows a peak at mass-ratios near unity (also seen in the higher-mass MS stars in 
the observed secondary-mass distribution), commonly referred to as ``twins'' and also observed in the Galactic field \citep[e.g.][]{rag10}. 
The model does not create such equal mass ratio systems, which suggests that these binaries are likely not formed dynamically and instead 
are a result of the binary formation process.

Recall that we have not yet derived the secondary-mass and mass-ratio distributions for the solar-type stars in M35.  Therefore 
these initial distributions are uncertain in our model.  
Investigating the cumulative distributions in Figures~\ref{M2fig} and~\ref{qfig} shows that the observable secondary-mass and mass-ratio 
distributions in the model have only changed slightly from their respective initial conditions.  
These results suggest that NGC 188 may have formed with
a secondary-mass distribution that was more strongly weighted towards lower masses (possibly following the IMF directly),
and also a mass-ratio distribution with a peak at mass ratios near unity.

Finally, we note that the period, eccentricity and secondary-mass distributions of the RGB binaries in the model
are all statistically indistinguishable from the respective MS distributions.
The model is consistent with the observed RGB period and eccentricity distributions, although we note that the small 
observed sample size prohibits a precise comparison.
(\citealt{gel12} did not derive the observed RGB secondary-mass or mass-ratio distributions for NGC 188.)

\subsection{The Role of Stellar Dynamics in Shaping the Main-Sequence Binary Population} \label{Sroledyn}

\begin{figure}[!t]
\plotone{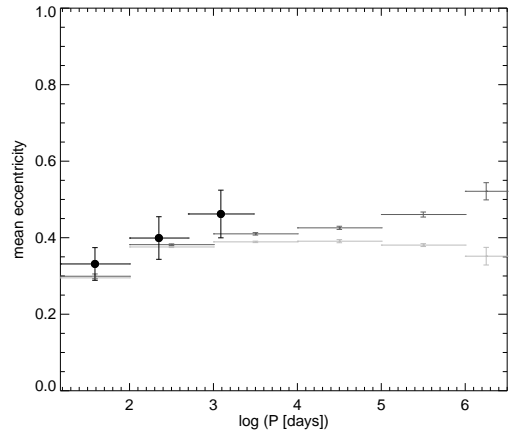}
\caption{\footnotesize
\label{evpfig}
Mean eccentricity as a function of orbital period for solar-type main-sequence -- main-sequence binaries in the model (with initial 
binaries in light gray and 7 Gyr binaries in dark gray) and observations (black).
We show binaries with $P > P_{circ}$, and plot the mean eccentricity over the bins in period shown in the horizontal 
lines.  Vertical error bars for the observed data show the standard errors of the mean.  For the model, we plot the 
mean values across all twenty simulations, and the vertical error bars show one standard deviation above and below these means.
At periods $>$ 1000 days, the 7 Gyr mean eccentricities become significantly larger than the initial eccentricities, due to 
stellar encounters.  A hint at a similar increase in eccentricity with increasing period is seen in the NGC 188 observations.
}
\epsscale{1.0}
\end{figure}

About 70\% of the MS-MS binaries present in the model at 7 Gyr maintained their primordial periods and eccentricities (both to within 10\%), 
despite 7 Gyr of dynamical evolution within the star cluster.  
Furthermore, 84\% of the 7 Gyr solar-type MS-MS binaries within our observable mass and period limits ($P < 3000$ days, $M_1 > 0.9$ \Msolar) 
maintained their primordial periods and eccentricities.
In short, the distributions of orbital parameters for the MS binaries, and especially for short-period binaries, were largely maintained throughout 
the cluster evolution.

In order to further investigate how dynamics has modified the binary population over the 7 Gyr of cluster evolution,
we compare the initial and 7 Gyr solar-type MS period, eccentricity, secondary-mass and mass-ratio distributions with those
derived by evolving the binaries from the model in isolation for 7 Gyr.
In the cumulative distribution panels of Figures~\ref{efig}~-~\ref{qfig}, the initial distributions are shown in the solid
lines.  The 7 Gyr solar-type MS distributions evolved within the cluster are shown in the light-gray (green in the online version) points, and the 7 Gyr
MS distributions evolved in isolation are shown in the dashed lines.

Given the union of twenty simulations, we are able to distinguish the initial and 7 Gyr MS binary period, eccentricity 
secondary-mass and mass-ratio distributions, respectively, all at very high confidence.  Likewise we can distinguish at
very high confidence these 7 Gyr binary distributions from the respective distributions derived from isolated evolution.  
However considering any individual simulation, only the 7 Gyr orbital period distribution has changed significantly from the 
initial distribution, with a significance level of $>$99\%.  Similarly the 7 Gyr MS period distribution evolved within the cluster
can be distinguished from that evolved in isolation at the $>$99\% confidence level for any given realization of the cluster.
All similar comparisons for the other binary distributions for any single simulation result in significance levels well below 90\%.

Only the longest period binaries experience significant orbital modifications from dynamical encounters.
In the case of the period distribution, we began our model with a fully populated log-normal distribution  (solid line Figure~\ref{Pfig}).  
Dynamical encounters have the most dramatic effect on the longest period, soft 
binaries, which are quickly disrupted by dynamical encounters, imposing a hard-soft boundary at $P \sim 10^{6.5}$ days.
The result is a period distribution at 7 Gyr that can be distinguished from the initial distribution at high confidence.
Conversely, the binaries evolved in isolation maintain the full log-normal period distribution and show no 
distinction from the initial distribution.

In the case of the eccentricity distribution, we began with a Gaussian distribution at all periods.  
Again, we see that dynamical encounters affect the longest period binaries most severely (Figure~\ref{evpfig}).
The eccentricity distributions for binaries with periods $>1000$ days are shifted to significantly higher eccentricities
than their initial
distributions, and the difference increases with increasing period.  Wider binaries have larger cross sections for encounters,
and such encounters (most often fly-by encounters) tend to increase eccentricity.  We note, however, that even binaries with periods 
very close to the hard-soft boundary do not attain a thermal eccentricity distribution in these simulations.

In Figure~\ref{evpfig} we also compare to observations of the NGC 188 MS binaries (black points).  We see a hint of a similar trend towards higher eccentricities
amongst longer period binaries in the observations.  However given the relatively small sample size, this possible trend in the observations is not
significant.  
As discussed in \citet{gel12}, many of these 
higher eccentricity binaries in NGC 188 reside near the cluster core, where dynamical encounters are most frequent.

Stellar dynamical and relaxation processes have also resulted in a modest increase in binary frequency (Figure~\ref{bfreqt}). 
The mean 7 Gyr solar-type MS hard-binary frequency over all simulations is 33.5\% (with a standard deviation of 2.8\%),
an increase of about 20\% over the initial MS solar-type hard-binary frequency.
This increase in binary frequency is due to the preferential evaporation of 
the lower-mass single stars as compared to the higher-total-mass binary stars in our observable 
mass range.  

As is clear from this analysis, the agreement between the 7 Gyr solar-type MS hard-binary population in the model with our observations
is largely due to our choice of initial conditions, as most  hard-binaries within our observable limit in orbital period
have not been dramatically affected by strong dynamical 
encounters over the 7 Gyr of evolution.  We return to this point in Section~\ref{discuss_M35NGC188}.

\section{Comparison Between the Observed and Simulated Blue Stragglers} \label{final_BS}

For the BS population in the model, we use an integrated sample that includes all BSs present at each $\sim$30 Myr snapshot interval within the twenty simulations 
from 6 to 7.5 Gyr, covering the range in the observed age estimates for NGC 188 \citep[e.g. see][]{for07}, 
and we will do similarly for the majority the paper.
The use of multiple simulations reduces the stochastic fluctuations in BS production.
The inclusion of all BSs at each snapshot interval weights the BS population by the lifetime that a given 
BS spends in each cluster radius, orbital configuration, etc., and essentially provides the likelihood that we will 
observe a BS in a given location or with a given orbital configuration, etc.

\citet{gel11} use the BS population from this model to examine the predicted binary orbital parameters of BSs formed through mass transfer
and stellar collisions and to investigate the origins of the long-period BS binaries in NGC 188.  
They conclude, based on the long-periods, low eccentricities and particularly the secondary-mass distribution which has a narrow peak at $\sim$0.5\Msolar, that the 
NGC 188 BSs in long-period binaries, which comprise the majority of the BS 
population in the cluster, likely have origins through mass-transfer processes. 
Here we expand upon their results.

First we examine the radial distribution of the BSs (Figure~\ref{cumR188}).
The observed NGC 188 BSs show a bimodal radial distribution with a centrally concentrated group of 14 BSs near the core and an additional 
population of 7 BSs in the halo \citep{gel08}.  \citet{gel12} find no significant difference between the binary properties of 
these two BS population in NGC 188, suggesting that this bimodal distribution is not a result of two distinct BS formation channels.
Similar bimodal BS radial distributions have been observed in globular clusters
\citep[e.g.][]{fer97} and have been attributed to dynamical friction and mass-segregation processes \citep{map04}.
These same processes act within our NGC 188 model; however the BS population in the model does not show the bimodal spatial 
distribution that we observe in the true cluster.
Moreover only 17\% of the simulated BSs are found outside of 5 pc ($\sim$3.5 core radii), as compared to $\sim$33\% in NGC 188.
Later we suggest that this difference is a result of incorrect modeling of BS formation channels (and particularly the mass-transfer channel)
which does not adequately population the halo with BSs.

\begin{figure}[!t]
\plotone{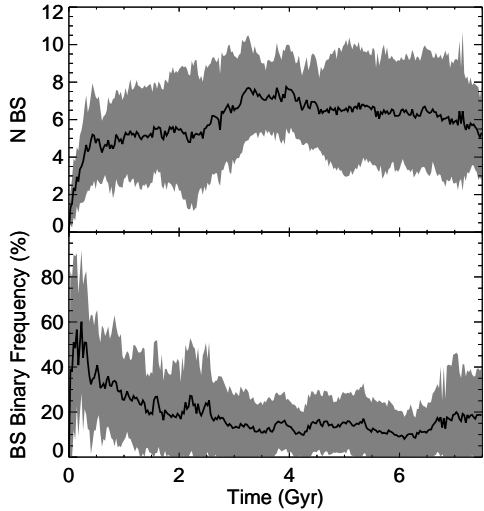}
\caption{\footnotesize
Number of blue stragglers (top) and the detectable blue-straggler binary frequency ($P<10^4$ days; bottom) as functions
of time.  In both panels, the black line shows the mean values at each snapshot interval found by 
averaging the results from all twenty simulations, and the gray region marks one standard deviation 
above and below the mean values.  In NGC 188, there are
21 observed blue stragglers, 16 of which are in hard binaries, resulting in a blue straggler hard-binary 
frequency of 76\%~$\pm$~19\%.  
In the model both the number of blue stragglers and the blue straggler binary frequency are significantly lower than the observed 
values.  
\label{NBSvtime}
}
\epsscale{1.0}
\end{figure}

In Figure~\ref{CMD} we show the 21 BSs observed in NGC 188 at 7 Gyr, 16 of which are in binaries with $P<10^4$ days \citep{mat09}.  
At 7 Gyr, the maximum number of BSs produced in
any of the twenty simulations is 11, and the mean number of BSs at 7 Gyr is $6.2 \pm 0.8$.
In Figure~\ref{NBSvtime} we show 
the number of BSs (top) and the BS hard-binary frequency ($P<10^4$ bottom) as functions of time for 
the NGC 188 model. 
We find on average $6.09 \pm 0.05$ BSs between 6 and 7.5 Gyr in our integrated BS sample from the model,
less than one third of the observed number of BSs in NGC 188.  

Next we compare the observed and simulated BS binary properties, starting with binary frequency.
The mean BS hard-binary frequency found in the model is also below the observed value of 76\%~$\pm$~19\%.
Between 6 and 7.5 Gyr, the mean detectable BS binary frequency is 14.5\%~$\pm$~0.5\%,
a factor of 5 lower than the BS hard-binary frequency observed in NGC 188.
As is clear from Figure~\ref{NBSvtime}, both the total number of BSs and the BS hard-binary frequency from the model 
are well below the observed values for essentially the entire lifetime of the modeled cluster.
Furthermore, the entire BS frequency, independent of period, between 
6 and 7.5 Gyr is only 27.0\%~$\pm$~0.4\%, about one third of the observed value (which only included BS binaries with $P < 10^4$ days).

We now return to Figures~\ref{efig}~-~\ref{qfig}, where we compare the distributions of the orbital parameters
for the integrated sample of BSs with those of the observed BSs in NGC 188.
Importantly, the forms of the distributions of orbital parameters for the detectable BS binaries in the model are 
consistent with those that we observe for the true BSs in NGC 188.
A large fraction of the simulated BSs are found in low-eccentricity orbits with long periods, and their 
secondary-mass distribution is peaked at $\sim$0.5~\Msolar.
Respective K-S test show that the BS binary period, eccentricity secondary-mass and mass-ratio distributions in the model can be distinguished from 
the associated solar-type MS binary distributions in the model at the $>$99\% confidence levels.  

This echos the results of \citet{mat09} and \citet{gel11,gel12} who compared the observed BS and solar-type MS binaries in NGC 188.
They find that the NGC 188 BSs in binaries have significantly longer periods than the MS binaries (at the $>$99\% confidence level), and that
amongst such long-period binaries, the BSs have significantly lower eccentricities (at the 98\% confidence level).
Furthermore, the companion-mass distribution for the long-period BS binaries in NGC 188 is narrowly peaked near 
$\sim$0.5~\Msolar, while the companion-mass distribution of solar-type MS binaries shows no evidence for a similar peak.
\citet{gel11} suggest that these BS binary properties point to an origin in mass-transfer processes, where the companions 
would be $\sim$0.5~\Msolar~white dwarfs.  

We note here that $\sim$60\% of the BSs in binaries with $P<3000$ days in the model 
derive from mass-transfer processes (and we expand upon this below).  Thus the agreement between the observed and simulated distributions of BS binary 
parameters is likely a result of both populations having the same dominant BS formation channel. 

In the model, we also find BSs in binaries at longer periods than our 3000 day completeness limit for observations (see Figure~\ref{Pfig}).
Currently five BSs in NGC 188 show no detectable radial-velocity variability, and are therefore considered ``single''.
However these NGC 188 BSs may prove to have companions in longer-period orbits (currently beyond our detection limit),
analogous to the longer-period BS binaries in the model.

Next, we analyze the BSs from the model according to formation mechanism.
BSs in the NGC 188 model form through three primary mechanisms, namely direct stellar collisions, mergers and mass-transfer processes.
Here we are concerned with the BS outcome.  Therefore in the following we will include within the merger category any binaries that are 
in the process of mass transfer and will eventually merge, which are predominantly Case A mass-transfer binaries between two MS stars.
Stellar collisions occur during dynamical encounters involving binaries (or higher-order systems), while mergers 
and mass transfer can occur within an isolated binary.  In the NGC 188 model
the rates of mergers and mass transfer are also modified by dynamical encounters (as we discuss in Section~\ref{discuss_BS}).  

Recently a new BS formation path was proposed \citep{iva08,per09}, where the inner binary in a triple system is driven to a merger 
by a combination of \citet{koz62} cycles and tidal friction known as the KCTF mechanism.  
This mechanism is available in the $N$-body model through the work of \citet{mar01} who added prescriptions to model
Kozai cycles and tidal friction in triple stars.  

In general, we find that binaries and triples contribute significantly to BS formation in the NGC 188 model.
The vast majority of the BSs formed in the model, regardless of the formation mechanism, had origins involving primordial 
binaries. Furthermore, nearly half of the BSs present in the NGC 188 model between 6 and 7.5 Gyr were found in dynamically formed 
hierarchical triple systems (in many cases with a primordial binary as the inner system) during the snapshot interval prior to becoming a BS.  
The wider orbit of the tertiary increases the cross section for stellar encounters, and in nearly all of these systems, the 
BSs formed through collisions resulting from dynamical interactions (as was predicted by \citealt{mat09} and \citealt{lei11}).  
Thus binaries and triples are key to BS formation in the NGC 188 model.

However, we do not find the KCTF mechanism to form a significant population of BSs in the NGC 188 model, and therefore we will not include
this mechanism in the following discussion of this section.  
Importantly, though, we did not include triples in the initial population of the model.
We return to BS formation through KCTF in Section~\ref{discuss_BS}.

\begin{figure}[!t]
\plotone{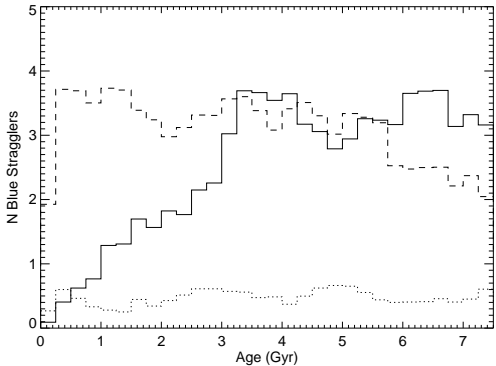}
\caption{\footnotesize
Number of blue stragglers as a function of time in the NGC 188 model for blue stragglers formed through collisions 
(solid line) mergers (dashed line) and mass transfer (dotted line), binned into 250 Myr intervals.  
Each bin is normalized to show the mean number of blue stragglers per simulation at a given time formed 
through the respective mechanisms.  Mergers dominate early in the cluster lifetime. 
At $\sim$3 Gyr collisions begin forming BSs at a rate 
equivalent to mergers. This corresponds with the time when the core begins to contract in the model.
The number of blue stragglers formed through mass transfer is always significantly
lower than those created through the other mechanisms. 
At the age of NGC 188 collisions and mergers dominate blue straggler formation in the $N$-body model.  
\label{NBSvtimevorg}
}
\epsscale{1.0}
\end{figure}

In Figure~\ref{NBSvtimevorg} we show the number of BSs as a function of time in the NGC 188 model, separated by formation mechanism, 
with BSs formed by collisions in the solid line, mergers 
in the dashed line and mass-transfer processes in the dotted line.  BS formation early in
the cluster lifetime is dominated by mergers of initially very short-period binaries. 
At $\sim$3 Gyr the
cluster core begins to contract, and we see a corresponding rise in the collision rate.  At the age of NGC 188
collisions are the dominant BS formation mechanism, followed closely by mergers.  The mass-transfer 
rate remains roughly constant throughout the lifetime of the cluster, and is in general
significantly lower than either the collision or merger formation rates.  

The relative contributions of the different BS formation mechanisms in the \citet{hur05} M67 $N$-body model are 
very similar to what we find here \citep[see][]{gel12}.
Of the 20 BSs present at 4 Gyr in the M67 simulation, 
eight formed through collisions, ten formed through mergers, and only two formed through mass-transfer processes.
As discussed in \citet{gel12} the M67 simulation contains a hard-binary frequency ($P < 10^4$ days) at 4 Gyr that is about twice that observed
in NGC 188 (or included in our NGC 188 model), and an overabundance of short-period binaries.
This large excess of binaries is responsible 
for the higher formation rate of BSs through all mechanisms in the \citet{hur05} simulations.

Next we compare the binary frequencies for BSs (within the integrated sample) formed by each formation mechanism in the NGC 188 model.
Only 4\% of the merger BSs in the NGC 188 model have binary companions with orbital periods below our
observational detection limit.  The majority of these are in contact systems that will eventually merge to form single BSs, while a 
small fraction have exchanged into detached binaries with wider separations.
Only 12\% of the BSs formed by collisions in the model would be detected as binaries.
Interestingly, even if we consider BS binaries at any period, only 33\% of BSs formed by collisions are found in binaries.  
Many of these single collisional BSs formed 
with binary companions initially in very wide orbits, near the hard-soft boundary, that were quickly ionized.
Thus the low observable-binary frequency for the BSs in the NGC 188 model reflects the dominant formation channels.  

Essentially all of the BSs formed by mass transfer in the NGC 188 model are in binaries that would be detectable observationally.
Only the mass-transfer mechanism can reproduce the large observed BS binary frequency of NGC 188.
Furthermore, about 50\% of the BSs that are detected as binaries in the NGC 188 model derive from mass-transfer processes.
Thus, although the absolute number of BSs formed by mass transfer in the NGC 188 model is low, mass transfer is 
the most efficient mechanism at producing BSs in binaries.
As discussed above, the agreement between the distributions of binary parameters for the BSs in the model with the BSs observed in NGC 188
is likely due to the majority of BSs in both samples having origins in mass-transfer processes.

Finally, we note that the NGC 188 model struggles to produce BSs with similar binary characteristics as the two double-lined (SB2) BSs in NGC 188.
Both of these SB2 BSs in NGC 188 have short periods ($P < 10$ days) and companions with masses $M_2>0.9$ \Msolar, one of which is a 
MS star near the turnoff and one of which is itself a BS.  
There are no BS binaries with $P < 10$ days and 
$M_2 > 0.9$ \Msolar~present 
in the model between 6 and 7.5 Gyr; there are two such systems present at earlier times.
Interestingly, out of the twenty simulations in the NGC 188 model, we find three BS-BS binaries present
between 6 and 7.5 Gyr with short enough periods to be detected as binaries observationally.
All of these systems formed through exchange encounters.
However these BS-BS binaries all have periods much greater than 10 days (specifically, 340, 2750, and 5500 days, respectively), and 
two of these three systems were only bound for one snapshot interval.  Another 4 BS-BS binaries are present in the integrated sample, but with 
periods $>10^4$ days, and therefore not detectable in our observations.  To date, no such long-period BS-BS binaries
have been detected in open clusters or the field.  

In summary, the deficiency in number of BSs, the low frequency of detectable binaries among those that are formed, and the 
lack of a bimodal BS radial distribution are 
striking failures of the model compared to the observations of \citet{gel11}. 
We return to this point in Section~\ref{discuss_BS}.
Importantly, however, the distributions of orbital periods, eccentricities, companion masses and mass ratios 
for the detectable BS binaries in the model (Figures~\ref{efig}-\ref{qfig})
are all consistent with the observed properties of the NGC 188 BSs.
The majority of these detectable BS binaries in the model were formed through mass-transfer processes. 
Thus, the agreement between the simulated and observed distributions of BS binary orbital parameters further supports the results of \citet{gel11},
who find that the majority of the true BSs in NGC 188 likely have origins in mass transfer.

\section{Dynamically Formed Triples} \label{final_trips}

Recently triple-star systems have been employed to explain a number of ``anomalous'' star systems 
observed in both star clusters and the Galactic field, including certain bright X-ray binaries \citep{mak09},
BSs \citep{iva08,per09}, and some short-period contact binaries \citep{egg06}.
As we did not include any triples initially, all triples discussed here were formed dynamically, 
often during binary-binary encounters.
Observationally, the triple populations in star clusters (including NGC 188) are poorly known.  Therefore here we mainly compare the triple 
frequency and distributions of orbital parameters in the NGC 188 model to those observed in the Galactic field, and 
investigate whether dynamical encounters can form triples with the same characteristics as those observed in the field.

\subsection{Frequency of Triples}

\begin{figure}[!t]
\plotone{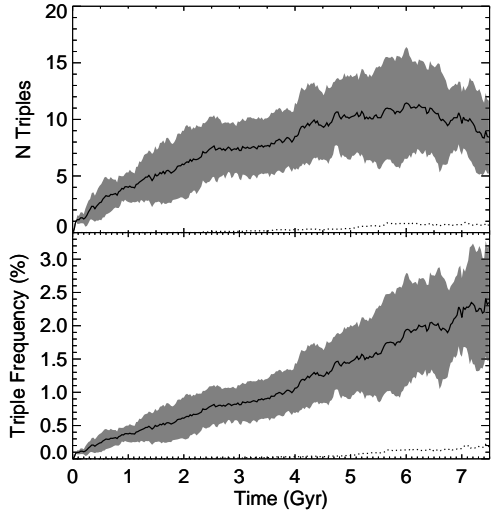}
\epsscale{1.0}
\caption{\footnotesize
Number of triples (top) and the triple frequency ($N_{trip}/N_{objects}$, bottom) as functions of time.
We include only triples with solar-type primary stars.
The black lines show the mean values at each snapshot interval for the twenty simulations, and 
the gray filled area shows one standard deviation above and below the means.
The dotted lines show the mean values for solar-type triples that have outer periods $<10^4$ days
(our spectroscopic detectability limit).
All triples formed dynamically, as we did not include triples initially in the model.
\label{Ntripvt}
}
\end{figure}

We first investigate the number of dynamically formed triples and the triple frequency 
($N_{trip}/N_{objects}$, where we treat each single star or system of stars as one object, respectively) as functions of time.
In Figure~\ref{Ntripvt}, the solid black lines show the mean number (top) and frequency (bottom) of triples with solar-type primaries ($0.9\leq M_1 $(\Msolar)$ \leq 1.1$)
in the model.  The dotted lines show the values for solar-type triples that we could potentially detect in our 
spectra, namely those with outer periods $<10^4$ days.
The number of solar-type triples increases early on due to initially very long-period binaries near (and beyond) the hard-soft boundary 
undergoing binary-binary encounters that result in dynamically formed triples,
as well as an increasing encounter rate in the core due to core contraction.
By $\sim$5 Gyr the number of solar-type triples reaches a maximum of $\sim$10 triples, and 
then declines as triples begin to evaporate from the cluster or are destroyed at a rate higher than they are created.
The average lifetime of a given triple in the model is $820 \pm 60$ Myr.

On the other hand, the solar-type triple frequency continues to increase throughout the entire lifetime of the modeled cluster.
These more massive systems are preferentially retained by the cluster as the less massive single stars
are ejected.  At 7 Gyr the solar-type triple frequency is 2.10\%~$\pm$~0.13\%.  However many of these triples have 
outer periods beyond our observational detection limit.

\citet{gel09} report two candidate triples 
(and one quadruple system) in their sample of binaries with orbital solutions in NGC 188, placing a lower limit 
on the MS triple frequency of 0.5\% (2/375), which certainly suffers from significant incompleteness.  
The NGC 188 model produces a detectable solar-type triple frequency of only 0.15\%~$\pm$~0.04\% at 7 Gyr, significantly lower than 
even this observed lower limit.

\citet{mer92} find a triple frequency 
of 2.3\%~$\pm$~1.6\% (2/88) for F5-K0 stars in the Pleiades ($\sim$150 Myr), and 
\citet{mer99} find F5-K0 stars in Praesepe ($\sim$750 Myr) to have a triple frequency of 3.8\%~$\pm$~2.2\% (3/80).
Both of these spectroscopic surveys have a roughly similar completeness limit as our NGC 188 survey.
The NGC 188 model produces essentially zero solar-type detectable triples at the ages of these young clusters (as is clear
from Figure~\ref{Ntripvt}).  If we take the full sample of solar-type triples (no longer limited to have $P < 10^4$ days),
we find a solar-type triple frequency of 0.11\%~$\pm$~0.01\% at 150 Myr and 0.31\%~$\pm$~0.03\% at 750 Myr, still well 
below the triple frequency detected observationally in these two clusters.
Thus at early times, the model does not form enough triples dynamically to match these observations.

Furthermore the triple frequency in the Galactic field is observed to be significantly higher than 
that found for the dynamically formed triples in the NGC 188 model.
\citet{rag10} find a triple frequency of 9\%~$\pm$~2\% in their complete sample of solar-type dwarf and subdwarf Galactic field stars.
This survey combined spectroscopic and astrometric techniques, and is therefore not limited by period; although we note that 
a fraction of these triples have outer periods beyond the hard-soft boundary in NGC 188.
The maximum triple frequency reached by any of the twenty simulations in the NGC 188 model is 4.5\% (at 7.2 Gyr).

\citet{tok06} performed a comprehensive survey of short-period ($1 < P < 30 $ days) 
solar-type Galactic field binaries within 100 pc, and find a tertiary ($N_{triple}/N_{binary}$) frequency of 65\%~$\pm$~5\%.
(Note that the denominator in this definition of the tertiary frequency is different from that of the triple frequency shown 
in Figure~\ref{Ntripvt} and discussed above.)  Accounting for triples in the \citet{tok06} sample that have outer periods beyond the hard-soft boundary of 
$10^{6.5}$ days reduces this observed frequency to $\sim$40\%.
However, the maximum tertiary frequency reached in the model for this mass and period range is only
10.9\%~$\pm$~1.7\% (at $\sim$7 Gyr), significantly lower than observed by \citet{tok06}.

\cite{tok06} also find the field tertiary frequency to be a strong function of the period of the inner binary, ranging from 
96\% ($1 < P $[days]$ < 3$) to 34\% ($12 < P $[days]$ < 30$).  For comparison, we find tertiary frequencies of 9\%~$\pm$~2\% and 10\%~$\pm$~3\%,
respectively, at 7 Gyr for these period bins. Thus
we do not find the same increase in the tertiary frequency for shorter-period inner binaries in our model.

\begin{figure*}[!t]
\centering
\epsscale{0.75}
\plotone{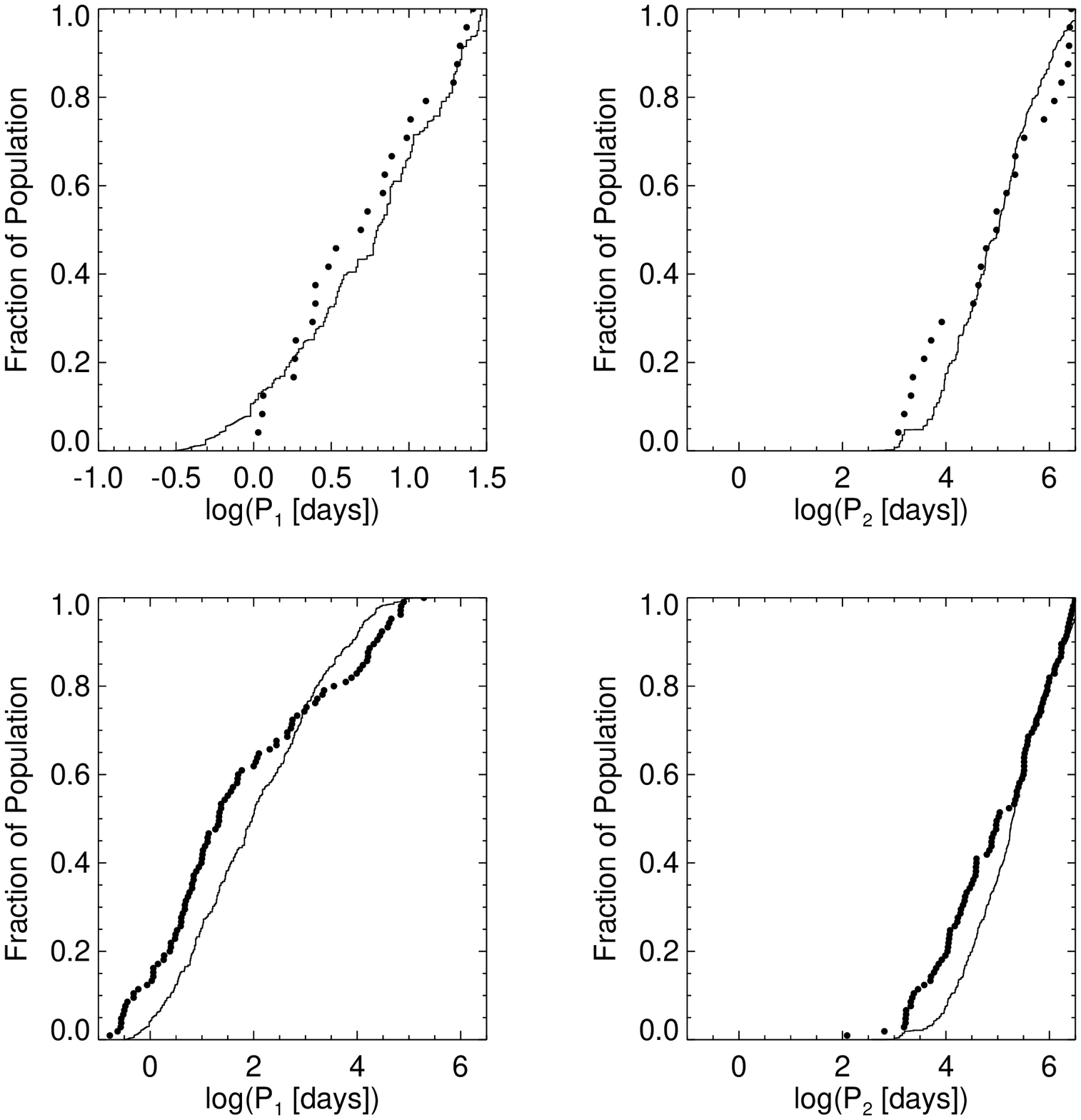}
\epsscale{1.0}
\caption{\footnotesize
Cumulative distributions of inner (left) and outer (right) orbital periods for solar-type dynamically formed triples in the NGC 188 model compared
to the observed field triple population of \citet{tok06} (top) and triples in the Multiple Star Catalogue  \citep[MSC;][bottom]{tok97}.
In all plots, the line shows the distributions from the model for triples containing solar-type (0.7~\Msolar~- 1.1~\Msolar) primaries 
present between 6 and 7.5 Gyr.  In the top panels, we limit the model sample to include only triples with inner-binary periods of $P_1<30$ days
(to correspond with the \citealt{tok06} study).
The black points in all plots show the observed samples including only triples with primary masses between 
0.7~\Msolar~and 1.1~\Msolar, and outer periods of $\log(P_2$ [days]) $< 6.5$ (the hard-soft boundary in the model between 6 and 7.5 Gyr).
Both the inner and outer orbital period distributions from the model are indistinguishable from those of the observed samples, respectively.
\label{Tokfig}
}
\end{figure*}

In conclusion, the frequency of dynamically formed triples produced within the NGC 188 model is significantly lower than triple frequencies
observed in the field and open clusters, suggesting that real open clusters may form with a significant population of primordial triples.

\subsection{Period Distributions of Dynamically Formed Triples}

In Figure~\ref{Tokfig} we compare the orbital periods for the solar-type triples in the NGC 188 model 
to those of the \citet{tok06} study (top)
and to those in the Multiple Star Calatogue \citep[MSC;][bottom]{tok97}.  
We limit both observational samples to include only triples with solar-type primary stars
and outer periods of $\log(P_2$ [days]$) < 6.5$ (the hard-soft boundary in the model at the age of NGC 188).
For comparison with the \citet{tok06} sample, we only include solar-type triples from the model with $1 < P_1 $(days)$ < 30$.
The MSC contains triple systems of all $P_1$, and therefore we do not limit the solar-type triple sample from the model 
for this comparison.

We note that the \citet{tok06} sample is nearly complete.  However, the MSC is a compilation of triples observed in various stellar environments 
from the literature (including those in \citealt{tok06}), and therefore is not meant to be a complete sample.  
Still this is the largest sample of triples in the literature, 
and therefore offers a useful comparison.

Remarkably, both the inner and outer orbital period distributions for dynamically formed triples in the model are indistinguishable 
from those of the similar triples in \cite{tok06} and the MSC, respectively.  Further observations are needed in order define
the distributions of other orbital parameters and component masses for a more in-depth comparison against $N$-body simulations.

\section{Discussion} \label{discuss}

The NGC 188 model matches the observed mass, surface-density profile, MS binary frequency and overall 
distributions of MS binary orbital elements of NGC 188 very closely. 
We discuss in Section~\ref{discuss_M35NGC188} the implications of this correspondence with NGC 188, given that our 
initial binary population was defined empirically based on our observations of M35.
On the other hand, the model creates a population of long-period circular MS-WD binaries not seen 
in NGC 188, and also creates too few BSs.  
Both of these highlight areas where the model requires improvement.
We discuss these issues in Sections~\ref{discuss_lPcirc} and~\ref{discuss_BS}, respectively.

\vspace{1em}
\subsection{Dynamical Evolution of the NGC 188 Solar-Type Binaries} \label{discuss_M35NGC188}

In Section~\ref{Sroledyn} we found that minimal changes to the solar-type MS binary population 
occurred in our NGC 188 model as a result of stellar dynamics, even after 7 Gyr of evolution within the star cluster.  
For the solar-type MS binary population, dynamical encounters only significantly affect the long-period binaries.
Most dramatically, strong dynamical encounters truncate the period distribution at the hard-soft boundary (Figure~\ref{Pfig}).
In Figure~\ref{evpfig} we also show that for binaries with $P > 1000$ days, stellar encounters begin to significantly shift the 
eccentricity distribution to higher eccentricities.  
Stellar dynamical and two-body relaxation processes also result in a modest increase in the global solar-type hard-binary frequency, and 
the central concentration of binaries relative to single stars.
However for any single simulation (e.g, any one realization of NGC 188) no distributions of binary orbital 
parameters or masses, expect for period, for the solar-type MS binaries at 7 Gyr show statistically 
significant changes as a result of dynamical processes.

These results underscore the importance of defining a realistic initial binary population in star cluster simulations, as
the properties of the initial hard-binary population persist throughout many Gyr of dynamical evolution (as was also 
discussed in \citealt{gel12}).  We suggest that this is best accomplished by comparison to observed binary populations.
Here we use the solar-type hard binaries from the young open cluster M35 as a guide, although all indications to date 
suggest that the field binaries are equally appropriate.

The NGC 188 model nearly reproduces the observed
NGC 188 MS solar-type hard-binary population at 7 Gyr.  This result
is due to the \textit{lack} of dynamical processing of the binaries, despite having evolved within a fairly massive open cluster for 7 Gyr.
The correspondence between the 7 Gyr MS solar-type hard binaries in the model and observations is
because NGC 188 and M35 themselves have very similar MS binary populations in both frequency and distributions of orbital parameters.

Our results suggest that, at the age of $\sim$180 Myr, NGC 188 likely had a very similar MS binary population as
M35, and indeed both M35 and NGC 188 may have formed with very similar binary populations.  The correspondence between 
the NGC 188 and M35 binaries with similar binaries in the field (which likely formed in a variety of different environments) 
introduces an intriguing possibility that these observed binary properties may be ubiquitous, at least for solar-type stars.
Observations of additional young open clusters, like M35, are essential to determine truly how common these binary properties 
are and the degree of scatter in the solar-type binary frequency and distributions of orbital parameters resulting from 
binary formation in different environments.  Given the results of the 
NGC 188 model, we suggest here that even observations of older open clusters can provide valuable information about the
properties of young binary populations.

\subsection{Long-Period Circular Solar-Type-Main-Sequence -- White-Dwarf Binaries} \label{discuss_lPcirc}

In Section~\ref{final_binary} we find a large population of solar-type-MS -- WD binaries in the model
that have periods well beyond the tidal circularization period, but with zero eccentricities (see Figures~\ref{elogPfig}~and~\ref{efig}).  
Such binaries are not observed in NGC 188 or any other solar-type binary population in the literature \citep[e.g.][]{gel12,mei05,duq91,rag10}.
Essentially all of these binaries in the model started with non-zero eccentricities, and were later circularized during 
a CE episode that began when the primary ascended the RGB or asymptotic-giant branch.
Similar binaries were also seen in the \citet{hur05} simulation, and were noted as an unrealistic population by \citet{gel12} in 
their comparison with NGC 188.
Here we examine two general questions in the hopes of resolving this discrepancy with observations: (1) Should 
these binaries truly go through CE evolution? And, (2) if they do go through CE evolution, should the products be MS - WD 
binaries in circular orbits with periods beyond the circularization period?

We begin by examining whether these binaries should truly go through CE evolution in reality.
In $\texttt{NBODY6}$, as in many binary evolution codes, the decision about whether to send a binary with one member (or both) overfilling
its Roche Lobe through dynamically unstable mass transfer (e.g., CE) of thermal/nuclear mass transfer is controlled by the critical mass ratio $q_c$.
Generally, in simulations utilizing $q_c$, a quantity $q_1 = M_{donor}/M_{accretor}$ is calculated, and binaries with $q_1 > q_c$ 
undergo CE evolution, while those with $q_1 < q_c$ undergo stable mass transfer.
Here we examine four different methods for calculating $q_c$ from the literature, namely those of \citet{hje87}, \citet{hur02},
\citet{che08} and the StarTrack code from \citet{bel08}.

The \citet{hur02} $q_c$ values are the lowest for any given binary in this sample, followed by the \citet{che08} criteria for conservative 
mass transfer, and then the \citet{hje87} values used in our NGC 188 model.  Thus these models would send all of these 
binaries through CE evolution.
\citet{bel08} use a slightly different procedure to determine when CE evolution will occur, which allows for a delayed dynamical 
instability for such binaries above a critical mass ratio $q_{ddi} = 3$ (defined similarly to $q_c$, above).
If we evolve these binaries with StarTrack, $\sim$37\% avoid CE, and $\sim$7\% undergo stable mass transfer
to become BSs.  

However, these $q_c$ values (including those used in the $N$-body model) are highly uncertain, and depend on how conservative the
mass transfer is \citep{che08}.
In the $N$-body code, conservative mass transfer is assumed when calculating the exponents for the boundaries between the 
dynamical, thermal and nuclear mass-transfer regimes (although non-conservative mass transfer is possible once the regime is determined).
We do not account for the effects of non-conservative mass transfer explicitly in the calculation of $q_c$.
However, \citet{woo12} argue that especially the initial stage of mass transfer from a 
giant is likely to be non-conservative due to high mass-loss rates. 
Furthermore, \citet{che08} show that non-conservative mass transfer effectively increases the $q_c$ value, which could allow 
additional binaries to avoid the CE phase, and potentially undergo stable mass transfer to produce BSs.

Also, enhanced wind mass loss during the primary's giant phase(s) will reduce the primary mass, and may also increase the 
secondary mass through wind accretion, both of which will reduce the $q_1$ value and potentially allow for stable mass transfer
in some systems.
In the $N$-body code, we use a Reimers mass loss coefficient of 0.5.
However the true Reimers coefficient for these stars may be as high as $\sim$1.4 \citep{kud78,sac93}, which would 
allow additional binaries to avoid the CE phase, and some to become BSs.

There is also growing evidence that this critical mass-ratio parameterization may be too simplistic \citep{woo11,pas11}.
Specifically, \citet{woo11} find that no parameterization based only on $q$ can reliably 
determine the onset of unstable mass transfer.
They also find that even for systems that will eventually undergo CE evolution, 
5\% - 10\% of the giant donor's mass may be transferred to the companion prior to this phase.
For many of the systems we discuss here, this amount of mass would be sufficient to produce a BS.

Furthermore, \citet{egg06a} shows from an observational sample of detached binaries possibly related to CE evolution that only those
with progenitor mass ratios $\gtrsim 4$ show the significant orbital shrinkage expected from a CE phase.
This may suggest that only systems with large mass ratios undergo typical CE evolution.
All the solar-type-MS -- WD binaries from the NGC 188 model discussed here have $q_1 \lesssim 2.5$.
Therefore perhaps most or even all of these binaries should avoid CE evolution in reality.

In summary, these unrealistic binaries produced in the $N$-body model may result from having incorrectly applied CE evolution to 
their progenitors.  In this scenario, these binaries could instead be BS-WD binaries at 7 Gyr with very similar orbital parameters 
and masses to the majority of the BSs observed in NGC 188.

Next we investigate the second question posed above, namely if CE evolution is truly expected for these binaries, should 
the products be MS-WD binaries in long-period circular orbits.  
The post-CE orbital periods in the model are determined by the uncertain $\alpha_{CE}$ efficiency parameter \citep[see][]{hur02}.
In our model (as in \citealt{hur05}) we attempt to mimic the treatment of \citet{ibe93} by setting $\alpha_{CE} = 3$.  
Doing so in the \texttt{NBODY6} treatment is approximately equivalent to using $\alpha_{CE} = 1$ in the \citet{ibe93} treatment,
as the two approaches use different equations for the orbital and binding energies.
These unrealistic MS-WD binaries may 
indicate that the $\alpha_{CE}$ value chosen here is incorrect. Perhaps the $\alpha_{CE}$ value (or values) for these binaries
should result in final periods that are less than $P_{circ}$ or even mergers (potentially producing BSs) after the CE phase.  

Finally, in the $N$-body model, we assume that all post-CE binaries are in circular orbits.  
However hydrodynamic simulations show that the envelope ejection can by asymmetric, which could impart a non-zero eccentricity to 
the system \citep{pas12}.
Additionally, material from the envelope may remain bound to the binary after the ejection phase \citep[e.g.][]{pas12,kas11,san98}.
This material may settle into a circumbinary disk, and such post-CE binaries can gain eccentricity through 
disk-binary interactions \citep{kas11,art91}.  In this scenario the product of the CE evolution may still be a long-period 
MS-WD binary, but with non-zero eccentricity.

To summarize, we identify three potential resolutions to this discrepancy with observations, which can be tested with future detailed 
simulations of these specific binaries.  First, perhaps the current MS members of these binaries should have gained sufficient mass to become
BSs by either undergoing stable mass transfer rather than CE (possibly by modifying the $q_c$ values) or by gaining mass prior to the CE phase.
Second, the $\alpha_{CE}$ value in the model may be incorrect and perhaps should be modified to shrink the orbits of these binaries 
shorter than $P_{circ}$ or even to result in mergers (possibly producing BSs) after the CE phase.  Third, if these binaries should indeed undergo CE 
evolution and produce long-period MS-WD binaries, perhaps they should have non-zero eccentricity.  

In closing, we note that there are roughly 9 such long-period circular binaries per simulation present at 7 Gyr.  If all were indeed supposed to become 
BSs through mass transfer this would more than double the 7 Gyr BS population in the NGC 188 model, and importantly, these BSs
would have similar binary characteristics to the majority of the NGC 188 BSs.  Thus it is intriguing to consider the possibility that 
both the paucity of BSs and the overabundance of long-period circular solar-type-MS -- WD binaries may be tied to our incomplete understanding 
of the same physical mechanism.
These binaries may provide important test cases for future detailed binary evolution simulations aimed at investigating CE evolution and 
mass-transfer processes.

\subsection{The Paucity of Blue Stragglers} \label{discuss_BS}

As discussed in Section~\ref{final_BS}, the NGC 188 model contains on average $\sim$6 BSs at the age of NGC 188
with a detectable binary frequency of $\sim$14.5\% ($P < 10^4$ days).  These values are both significantly lower than 
those observed in NGC 188, where we find 21 BSs with a binary frequency of 76\%~$\pm$~19\% (within the same period range).
Here we identify potential reasons for the paucity of BSs in the NGC 188 model, and suggest future work that may further 
illuminate how BSs form in open clusters.

\subsubsection{Low Mass Transfer Efficiency} \label{lowMT}

\citet{gel11} find that the binary properties of the long-period
NGC 188 BSs (and particularly the companion-mass distribution) are most closely consistent with an origin in mass transfer.
We find above that only the mass-transfer mechanism can reproduce the very large hard-binary frequency and distinctive 
distributions of orbital periods and companion masses observed for the NGC 188 BSs.

However, the mass-transfer mechanism is the least efficient at producing BSs in the NGC 188 model.  
If we evolve the primordial 
binaries from our twenty simulations in isolation (using the \citealt{hur02} BSE code), 
we would expect that at 7 Gyr on average $\sim$5 BSs per simulation would
have formed through mass transfer (not accounting for the expected mass loss from the cluster due to 
tidal stripping and dynamical ejections).
Therefore the efficiency of the mass-transfer formation mechanism is far too low in the $N$-body model than 
would be necessary to produce the $\sim$13 BSs in NGC 188 that we expect are the result of mass-transfer processes.

In fact, these five mass-transfer BSs that results from evolving the binary population from the model in isolation are lost in 
a dynamical environment.  Dynamical encounters disrupted, modified and/or ejected nearly all of these ``proto-BS'' binaries in the NGC 188 model.
On the other hand dynamical encounters do not efficiently create BSs from mass-transfer processes within the 
NGC 188 model.  Only on average $\sim$0.4 additional BSs per simulation were formed via mass transfer that were not expected 
from the population synthesis analysis.

\begin{figure}[!t]
\centering
\plotone{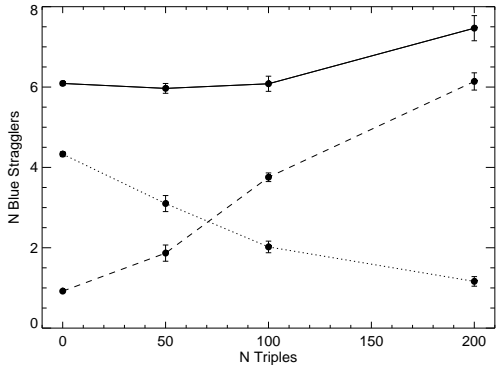}
\epsscale{1.0}
\caption{\footnotesize
Mean number of blue stragglers between 6 and 7.5 Gyr as a function of the number of primordial triples in the $N$-body simulations.
The values at zero primordial triples are from the NGC 188 model.  All other values are from the additional simulations described in 
Section~\ref{primtrips}.  The points connected by the solid line show the mean total number of BSs.  The points connected by the 
dashed line show the mean total number of hard-binary BSs ($P < 10^4$ days), and the points connected by the dotted line show the 
mean total number of single BSs.  We find here that the total number of BSs increases 
only slightly with an increased number of primordial triples, but the hard-binary frequency among the BSs increases dramatically.  Still the 
total number of blue stragglers in these simulations are a significantly lower than the number (21) observed in NGC 188.
\label{NBSvtripfig}
}
\end{figure}

In Section~\ref{final_BS} we note that the NGC 188 model lacks the bimodal BS radial distribution of the true cluster, and 
instead the BSs in the model are centrally concentrated, with no significant halo population.  This discrepancy between the 
model and observations may be tied to the inefficient production of BSs through mass transfer, where, in the model, insufficient BSs are 
produced in the halo to yield a bimodal distribution.
Additional $N$-body models are necessary to fully understand the implications of 
this discrepancy between observations and simulations.  

We suggest that the paucity of 
BSs in the NGC 188 model reflects upon an incomplete description of the mass-transfer mechanism in the model. The presence of
a population of unrealistic solar-type-MS -- WD binaries with long-periods and circular orbits further highlights our incomplete understanding of 
the interface between CE evolution and stable mass transfer.  As discussed in Section~\ref{discuss_lPcirc}, if these spurious MS-WD binaries 
were instead meant to undergo stable mass transfer to become BSs, the NGC 188 model would roughly match the observed BS number and binary properties.
Our analysis of the BSs produced in the NGC 188 model combined with the results 
of \citet{gel11} suggest that the efficiency of stable mass transfer may be significantly underestimated in the $N$-body model (and other codes
that use similar binary evolution prescriptions).

\subsubsection{Effects of Primordial Triples on Blue Straggler Production} \label{primtrips}

Another hypothesis is that the lack of BSs in the model is the result of a missing population of hierarchical triples that could 
produce BSs either through mergers of the inner binaries as a result of Kozai cycles and tidal friction \citep{iva08,per09},
known as the KCTF mechanism, or through collisions during dynamical encounters \citep{mat09,lei11}.
Kozai cycles and tidal friction are included in the $N$-body model as in \citet{mar01}. Stellar encounters involving triples that may result 
in collisions are also modeled in detail.

We did not include triples in the primordial population of the NGC 188 model, but triples do form dynamically (see Section~\ref{final_trips}).  
In Section~\ref{final_BS} we find that many of the BSs in the NGC 188 model that formed through collisions were 
members of dynamically formed hierarchical triples prior to the encounter that resulted in the collision.
This is fully consistent with the analytic work of \citet{lei11}, who show that dynamical encounters involving triples may be common in clusters like NGC 188, 
and can lead to BS formation through collisions.  

We also investigate the NGC 188 model for dynamically formed triples that have appropriate orbital parameters to potentially form BSs through 
the KCTF mechanism.  Specifically, we search for triples with inner binaries that have a combined mass $>$1.15~\Msolar~(roughly 
the lower mass limit of BSs in NGC 188), inner periods of $<6$ days (as in \citealt{per09}), and outer periods 
between 700 and 3000 days (roughly the period range of the long-period NGC 188 BSs).  On average, we find $\sim$1 such triple 
per simulation at any given time, and no BSs present in our model between 6 and 7.5 Gyr formed through the KCTF mechanism.
Only three BSs present at other times within our twenty simulations formed as a result of Kozai cycles.

However, it may be common for short-period ($P \lesssim 6$ days) binaries to form with tertiary companions \citep{tok06,fab07},
which are conducive to forming BSs through the KCTF mechanism and collisions resulting from dynamical encounters.
Furthermore, in Section~\ref{final_trips}, we find that the triple frequency in the NGC 188 model is lower than expected from observations
of real open clusters (and the Galactic field), suggesting that perhaps triples should be included in the initial population.

As a first step toward investigating the importance of primordial triples for BS production, we generated a few additional simulations using the 
same initial conditions as in the NGC 188 model, but also including up to 200 primordial triples.  
For simplicity and to emphasize the types of triples that may be important for producing BSs in a cluster like NGC 188 through the KCTF mechanism, 
in these simulations we choose the primaries in all initial triples to have
masses of 1 \Msolar, the secondaries to have masses chosen randomly from a uniform distribution between 0.2~\Msolar~and 1 \Msolar, and 
the tertiaries to have masses chosen randomly from a uniform distribution between 0.5~\Msolar~and 2.0~\Msolar.
We choose the inner periods from the \citet{duq91} log-normal distribution (consistent with the M35 solar-type binaries), but limited 
to be between 2 and 50 days. 
The cut in inner period at 50 days is motivated by the observations of Galactic field triples by \citet{tok06} who find that the tertiary frequency 
falls off quickly with increasing inner binary period.
We impose a lower limit on the inner period of 2 days so that we don't introduce binaries that will quickly merge on their own through Case A mass transfer 
(which occurs at periods $\lesssim1.7$ days for solar-type stars in BSE).
The outer periods are chosen randomly between 700 and 3000 days, to reflect the periods of the observed NGC 188 single-lined (SB1) BS binaries. 
The eccentricities for both orbits are chosen from the M35 Gaussian distribution. All angles are chosen randomly.

For these simulations, we modify the initial frequency and period distribution of the regular (non-triple) binary population so that when the two components 
of the triple systems are combined with the regular binary population, the simulations still begin with the same log-normal period 
distribution and the initial binary frequency used in the NGC 188 model. There are $\sim$200 solar-type MS binaries in the initial population 
of the NGC 188 model with $2 < P($days$)< 50$, which sets the maximum number of triples that we include in these simulations.

We ran simulations with 50, 100 and 200 primordial triples, and compare the results to our NGC 188 model.  In Figure~\ref{NBSvtripfig} 
we show the mean number of BSs present in these simulations between 6 and 7.5 Gyr.  The points at zero triples come from our ``standard'' 
NGC 188 model without primordial triples.  The points connected by the solid line show the mean total number of BSs in these simulations.  
There is a slight increase in the number of BSs with increasing number of triples, but even in the simulation containing 200 primordial triples
(essentially replacing every solar-type binary with $2 < P$[days]$< 50$ in the standard NGC 188 model with a triple) 
the simulation still only creates about one-third of the number of BSs that are observed in NGC 188.

The BS hard-binary frequency ($P < 10^4$ days) increases dramatically with increasing number of triples, as more BSs are created 
within triples, and therefore retain companions.  Also note that this increase in BS hard-binary frequency is in part due to setting the 
initial outer orbits to have periods between 700 and 3000 days, which are not easily disrupted.  
About half of the BSs in binaries that are present in these simulations 
between 6 and 7.5 Gyr have origins from primordial triples.  
The majority of these BSs that began within primordial triples were involved in dynamical encounters that 
perturbed the system and in many cases exchanged in other stars prior to a collision that formed the BSs.
Only about 25\% of these formed in relative isolation through Kozai-induced BS formation.  
Thus, triples do play a significant role in creating BSs through stellar collisions (as was suggested by \citealt{mat09} and \citealt{lei11}),
but the KCTF mechanism is not a significant contributor to the BS population here.

The \texttt{NBODY6} code uses the \citet{mar01} analytic method to model Kozai oscillations and tidal processes in triple stars, 
rather than direct integration.  This method is used because direct integration of triple (and higher-order) stars 
for many Myr within a cluster simulation is computationally very challenging and could also introduce further uncertainties 
due to the many necessary integration steps, each with its own computational uncertainty.
It is possible that the \citet{mar01} method underestimates the efficiency of KCTF BS formation, which would affect the results shown here.  
Further investigation into 
the efficacy of the \citet{mar01} method for KCTF BS formation in $N$-body simulations is desirable, but beyond the scope of this project.
Additionally, further $N$-body simulations with more realistic primordial triple populations are required to fully test this hypothesis.

In summary, our results here suggest that primordial triples may not provide the additional BSs that are required to bring the number of BSs in 
the NGC 188 model into agreement with those observed in the true cluster.  
Specifically, even replacing all solar-type binaries with $2 < P$(days)$< 50$ in the standard NGC 188 model with triples 
does not reproduce the number of 
BSs observed in NGC 188.  The majority of the BSs produced in these models through triple-mediated mechanisms were produced by collisions.
However \citet{gel11} rule out collisions as the dominant formation mechanism for the majority of the NGC 188 BSs at high confidence (based primarily
on the observed secondary-mass and orbital eccentricity distributions of the NGC 188 BSs).

\subsubsection{Dependence of Blue Straggler Production on the Initial Binary Population and Cluster Structure} \label{primbins}

The BS production rate is sensitive to the initial binary parameters, and certain modifications can increase the 
number of model BSs to be consistent with the 21 observed in NGC 188.  We describe the results of additional $N$-body simulation 
with modified initial conditions below.  However none can reproduce the observed binary properties of the NGC 188 BSs. 

Adding short-period binaries (e.g., by using an initial period distribution that is flat in log period) is the most efficient method 
for increasing the number of BSs.
Here the BS production rate through mergers and collisions are increased, as was also seen in the \citet{hur05} simulation.
However as discussed above such BSs do not match the binary properties of the NGC 188 BSs.
Adding more high-eccentricity binaries (e.g. by using an initially thermal eccentricity distribution)  increases the BS production
rate through collisions, as small perturbations to a highly eccentric orbit can cause the components of the binary to collide.  
Dramatically increasing the overall hard-binary frequency can also increase the BS production rate through all mechanisms.
However, modifying the initial period or eccentricity distributions or the binary frequency in this manner results in a 7 Gyr MS binary 
population that is inconsistent with that observed for NGC 188.

\citet{gel12} find that the NGC 188 MS binaries have a modest abundance of ``twin'' binaries which is not reproduced in 
our model.  Adding a population of twins to the initial binary population does not significantly increase the BS production rate.

We also ran additional $N$-body simulations to investigate the effect of changing the initial half-mass radius of the cluster 
(between 4 and 7 pc) and find no significant effect on 
the BS production rate at the age of NGC 188.  After 3 - 5 Gyr of evolution, the clusters relax to 
similar core radii regardless of the initial values, and by 7 Gyr the core (and tidal) radii in all of these additional simulations are consistent with 
that of the NGC 188 model.  Since the typical BS lifetime in the model is $\sim$1.6 Gyr, the difference in cluster structure 
at a young age does not significantly affect the BS population at the age of NGC 188.

In summary, we are unable to modify the initial conditions of the binary population or the cluster structure
in such a way as to both reproduce the observed BS and MS binary populations.

\subsubsection{Blue Straggler Lifetimes} \label{BSlifetimes}

It is possible that the paucity of BSs in the model are a result of underestimated BS lifetimes.
Numerous uncertainties, including, for example, how efficiently fresh Hydrogen can be mixed into the core, translate into
uncertainties on the BS lifetimes in the NGC 188 model.   
For BSs present at 7 Gyr in the NGC 188 model, the 
mean BS lifetime is $\sim$1650 Myr (not including prior evolution as a normal MS star).  
In order to match the number of BSs observed in NGC 188, we would have to approximately triple the lifetime of all BSs. 
However the total lifetime of a normal MS star with a mass of a typical 7 Gyr BS in the model ($\sim$1.46~\Msolar), is 
less than twice that of the mean BS lifetime.  Therefore tripling all BS lifetimes would be unphysical.
Furthermore, \citet{gle08} suggest that BSs, and particularly those formed by collisions, will be out of thermal equilibrium and poorly mixed, and 
therefore will have \textit{shorter} lifetimes than predicted by the equilibrium models used in the $N$-body code.
Finally, simply increasing the lifetime of all BSs would likely not change the BS binary frequency.

\section{Summary and Conclusions} \label{conclusion}

In this paper we present a sophisticated $N$-body model of the old (7 Gyr) open cluster NGC 188 that 
matches the observed MS and RGB binary populations in detail, and thereby allows us to study the formation rate and mechanisms of 
the BS population within a realistic theoretical framework.
We take great care to define the initial conditions for the model empirically, where possible.
Importantly, we employ our observations of the solar-type binary population of the young open cluster M35 (180 Myr) to 
guide our choices for the initial binary frequency and distributions of period and eccentricity (and we note that these M35 binary 
characteristics are consistent with those of solar-type binaries in the Galactic field).

The power of using detailed observations to both guide the choices for initial conditions and test the outcomes of $N$-body open 
cluster simulations
is evident in the accuracy with which we reproduce the observations of the NGC 188 solar-type binaries with our model.
The MS binary frequency and distributions of orbital parameters agree in detail with 
those observed in the real cluster \citep{gel12}.  
Additionally, at the age of NGC 188 the model matches the observed cluster mass as well as the core and tidal radii of the 
cluster \citep{bon05,gel08,chu10}.  

The MS binaries show very little evidence for modifications by stellar dynamical encounters even after 7 Gyr of evolution within the cluster environment.
Indeed only the long-period binaries show evidence for orbit modifications from stellar encounters.
Specifically, strong dynamical interactions are responsible for breaking up the initially very long-period binaries, and establishing a hard-soft 
boundary at $P \sim 10^{6.5}$ days.
Stellar encounter (most often fly-by encounters) begin to shift the eccentricity distribution for MS-MS binaries to higher eccentricities 
for binaries with $P > 1000$ days, and this effect increases towards longer periods.
Two-body relaxation processes also result in a $\sim$20\% increase in the detectable binary frequency over the 7 Gyr of evolution and the central 
concentration of the MS binaries in the cluster at the age of NGC 188.
However, in general, the short-period MS binaries in the NGC 188 model show little signs of dynamical processing, which further 
emphasizes the importance of defining an accurate initial binary population for constructing realistic open cluster models, 
as most of the hard-binary characteristics persist throughout the evolution of the cluster.

This finding may also suggest that observations of the present-day binaries in open clusters can provide valuable 
information on their primordial binary populations.  The similarities between the solar-type binary populations in NGC 188 (at 7 Gyr), M35
(at 180 Myr) and the Galactic field are remarkable, and may indicate that binaries form with similar characteristics within a variety of environments.
Comparable studies of binary populations in additional rich open clusters are necessary to explore this further.
Nevertheless, we suggest that the observationally defined distributions of binary orbital parameters and binary frequency used here should be strongly 
considered for use in future open cluster models.

However, despite having matched the observed MS binaries, the NGC 188 model produces only one-third as many BSs, and 
these BSs have only one-fifth the hard-binary frequency ($P < 10^4$ days), as compared to the BSs observed in NGC 188.  
For the full 7 Gyr of the modeled cluster, mergers and collisions dominate
BS formation, with collisions producing the majority of the BSs at the age of NGC 188.  This is in stark contrast to the 
observations of \citet{gel11} who show that the secondary-mass distribution of the majority of the BSs in NGC 188 is inconsistent 
with the collisionally produced BSs in the model.  Rather the binary properties of the NGC 188 BSs, and particularly the secondary-mass distribution, 
point to a mass-transfer origin.

A potential explanation for the missing BSs may also resolve a second discrepancy between the model and observations, 
namely the population of circular MS-WD binaries with periods well beyond the tidal circularization period created within the model as a result of
CE evolution.  Such binaries are not observed in NGC 188, the Galactic field \citep{duq91,rag10} or the many other open cluster surveyed 
by \citet{mei05}.  
Perhaps the MS stars in these binaries were instead meant to become BSs by entering a phase of stable mass transfer.
This is a compelling solution as the number of unphysical MS-WD binaries in the model is consistent with the number of 
missing BSs, and the binary characteristics of these BSs would be consistent with the majority of the NGC 188 BSs. 

The creation of these unphysical long-period circular MS-WD binaries in the model highlights our incomplete understanding of CE evolution,
and particularly which binaries should enter a CE phase, which should enter stable mass transfer and what the products of CE evolution should look like.
Improvements may be necessary in the treatment of mass transfer and CE evolution in models like \texttt{NBODY6} and BSE,
which may also have important implications for formation rates of other interacting binary populations,
such as W UMa's, CVs, symbiotic stars, etc.
These specific unphysical binaries created in the NGC 188 model may provide important test cases for future detailed binary evolution 
simulations aimed at studying CE evolution and mass-transfer processes.

Missing BSs in the NGC 188 model may also point to a missing population of primordial triples. The Pleiades \citep{mer92} 
and Praesepe \citep{mer99} both have significantly higher observed triple frequencies than the NGC 188 model at the respective cluster ages,
as does the Galactic field \citep{duq91,tok06,rag10}.  However, our additional simulations (Section~\ref{primtrips}) suggest that 
adding primordial triples will not dramatically increase the number of BSs present at 7 Gyr.
Still, additional $N$-body simulations, with more realistic initial triple populations, are required to fully investigate if primordial triples 
can contribute significantly to BS production.

In closing, we note that only recently have both the observations and $N$-body simulations matured sufficiently to permit such detailed comparisons as we present here
between true binaries in a dynamically evolved open cluster and analogous binaries simulated in a realistic $N$-body open cluster model.
Open clusters are unique in their accessibility to both detailed comprehensive observational surveys and full-scale sophisticated $N$-body simulations, 
opening a window into the dynamical evolution of star clusters that is only now becoming reachable.  Furthermore, advances in both hardware and software now 
readily allow for multiple large open cluster simulations to run simultaneously (as we have done here), which enable more statistically 
robust analyses and a broader exploration of parameter space \citep{mat08}.  This capability, combined with the upcoming results 
from our WOCS observations of a number of open clusters with a wide range in age, will continue to advance our understanding of star cluster 
dynamics, the origins of BSs and the roles that binary and multiple stars play in star cluster evolution.

\acknowledgments
Thanks to R.~Taam and N.~Ivanova for helpful discussions.
This work was funded by the Lindheimer Fellowship at Northwestern University, National Science Foundation (NSF) East Asia and Pacific 
Summer Institute (EAPSI) award OISE-0913544, NSF grant AST-0908082 to the University of Wisconsin - Madison, and the Wisconsin Space Grant Consortium.

\bibliographystyle{apj}                       
\bibliography{ngc188.nbody.ms}

\end{document}